\documentclass[preprint2]{proto}
\usepackage{times}
\newcommand{\refs}{\par\noindent\hangindent=1pc\hangafter=1}
\pdfoutput=1
\begin{document}

\title{\textbf{\LARGE The Link between Magnetic Fields and Cloud/Star Formation}}

\author {\textbf{\large Hua-bai Li}}
\affil{\small\em Max Planck Institute for Astronomy; The Chinese University of Hong Kong}

\author {\textbf{\large Alyssa Goodman; T. K. Sridharan}}
\affil{\small\em Harvard-Smithsonian Center for Astrophysics}

\author {\textbf{\large Martin Houde}}
\affil{\small\em University of Western Ontario; California Institute of Technology}

\author {\textbf{\large Zhi-Yun Li}}
\affil{\small\em University of Virginia}

\author {\textbf{\large Giles Novak}}
\affil{\small\em Northwestern University}

\author {\textbf{\large  Kwok Sun Tang}}
\affil{\small\em The Chinese University of Hong Kong}


\begin{abstract}
\baselineskip = 11pt
\leftskip = 0.65in 
\rightskip = 0.65in
\parindent=1pc
{\small The question whether magnetic fields play an important role in the 
processes of molecular cloud and star formation has been debated for 
decades.  Recent observations have revealed a simple picture that may help 
illuminate these questions: magnetic fields have a tendency to preserve 
their orientation at all scales that have been probed - from 100-pc 
scale inter-cloud media down to sub-pc scale cloud cores.  This ordered 
morphology has implications for the way in which self-gravity and 
turbulence interact with magnetic fields: both gravitational 
contraction and turbulent velocities should be anisotropic, due to the 
influence of dynamically important magnetic fields.  Such anisotropy is now 
observed.  Here we review these recent observations and discuss how they 
can improve our understanding of cloud/star formation.
 \\~\\~\\~}

\end{abstract}

\section{\textbf{INTRODUCTION}}

How molecular clouds form and then fragment into filaments, cores, protostellar discs and finally stars is still mysterious. The existence of large-scale magnetic fields (B-fields) in the interstellar medium (ISM) was first shown by {\em Hiltner (1951)} and {\em Hall (1951)} nearly
65 years ago.  On the cloud-formation scale ($10^{3}-10^{2}$ pc), B-field strength is expected to determine whether a cloud can rotate with the angular momentum inherited from galactic shear and/or turbulence. The fragmentation of a cloud (e.g., the formation of filamentary structures, see Chapter by Andr\'e et al. in this volume) also strongly depends on the cloud B-field strength. For protostellar disc formation, the debate is centered on whether {\em magnetic braking} can prevent the formation of large rotationally supported discs at early times (see Chapter by {\em Zhi-Yun Li et al.}). At almost all stages of cloud/star formation, the roles played by B-fields remain highly controversial, in large part as a result of the lack of observational constraints.

This situation is gradually improving, owing to data obtained in some decade-long surveys of B-field strength (Zeeman measurements; e.g., {\em Troland and Crutcher},  2008) or morphologies (sub-mm polarimetry from, e.g., {\em Dotson et al.} 2010 and {\em Matthews et al.} 2009; IR polarimetry from, e.g., {\em Clemens et al.} 2012) of molecular clouds having been released recently. In this paper, we review a series of relatively recent {\em surveys} (23 surveys are selected and divided into 14 categories; Figure 1) that lead us to a picture where magnetic fields play a variety of roles at different points during the star formation process.  A significant portion of this chapter is devoted to the {\em magnetic topology problem} raised in PPIII ({\em McKee et al.} 1993): how does the B-field topology evolve as molecular clouds form out of the interstellar media (ISM) and as cores within the cloud contract to form stars? After 20 years, recent surveys have finally shed some light on this problem. 

We have organized the following discussion by the scales involved in each process: from section 2 to 5, respectively, cloud formation, filament formation, core formation and protostellar disc formation. In section 6, we synthesize the results from various observations and try to explain their discrepancy. 

\bigskip 
 \begin{figure*}[float]
  \includegraphics[trim=.4cm .2cm 0cm 0cm, clip=true, totalheight=0.28 \textheight, angle=0]{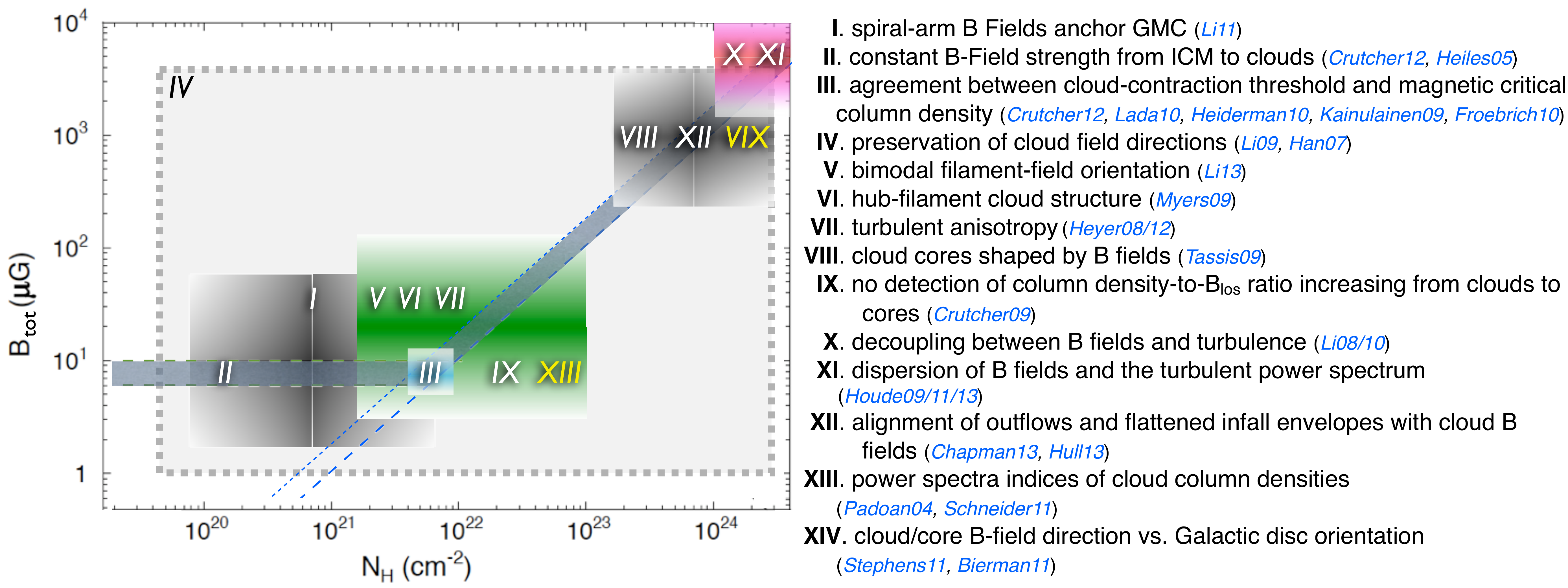}
 \caption{\small 23 recent surveys (noted by the leading authors and years) are discussed here in 14 categories. The rectangles in the plot show the limits of the typical B-field strengths and column densities the observations probed.} 
\end{figure*}

\begin{figure*}
 \epsscale{2.}
 \plotone{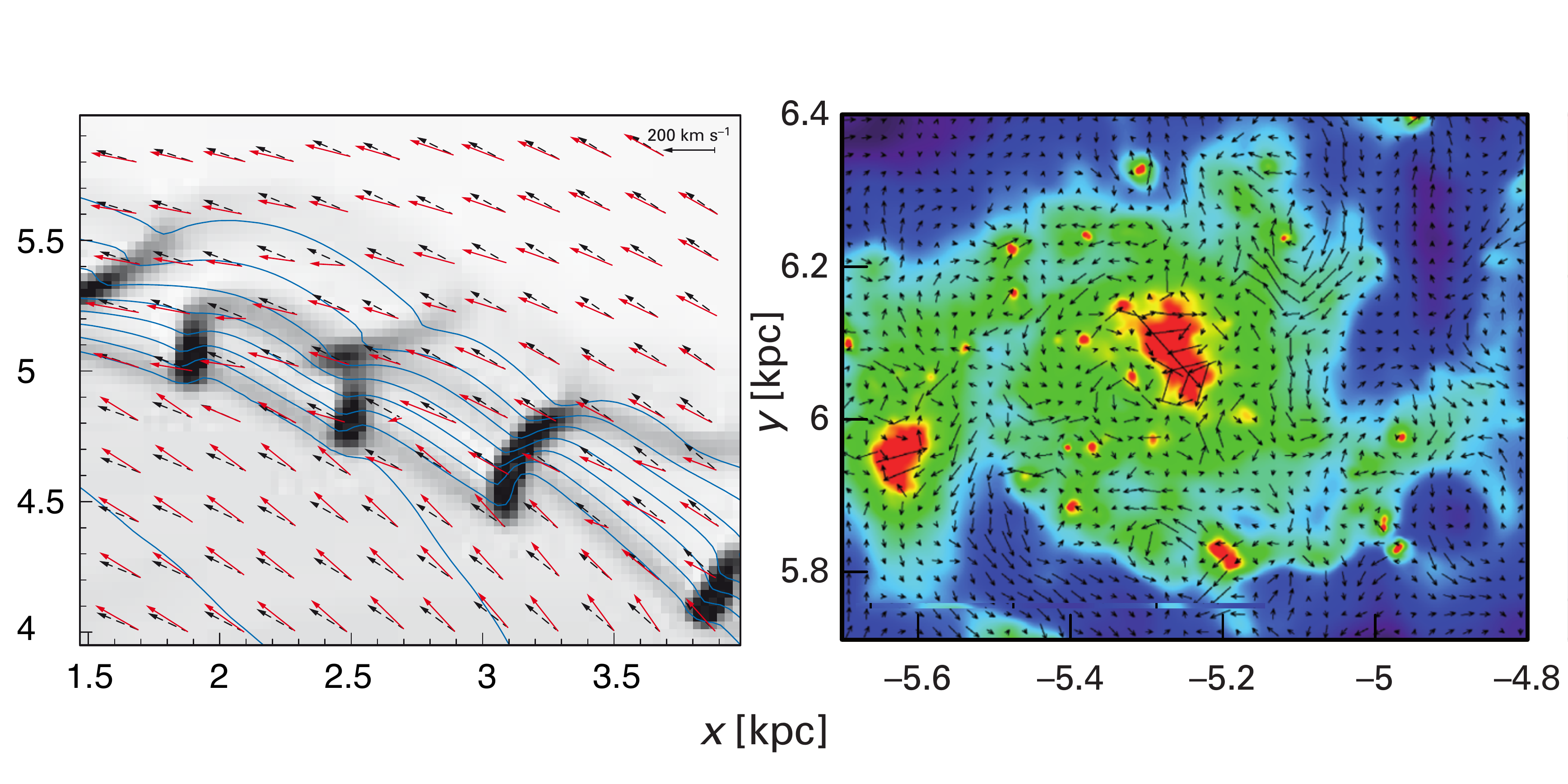}
 \caption{\small Two competing scenarios of cloud formation. Left: A patch from a global galaxy simulation ({\em Shetty and Ostriker} 2006). The solid vectors show the instantaneous gas velocity in the frame rotating with the spiral potential. The dotted vectors show the initial velocities (pure circular motion). The solid blue lines show B-field orientations. The gray scale stands for the relative surface density. The B-fields of the spiral arm are only slightly twisted in the molecular cloud complexes (dark elongated regions), and in turn the field tension is strong enough to hinder the cloud rotation. Right: A similar simulation ({\em Dobbs et al.} 2008) but the well developed cloud rotation has produced tidal tails extending from the GMC, and the B-fields (vectors) follow the rotation and have lost the ''memory'' of the galactic field direction. For color figure see the arXiv version.
}  
 \end{figure*}

\bigskip  

\centerline{\textbf{ 2. MOLECULAR CLOUD FORMATION }}
\bigskip

The formation of molecular clouds is poorly understood. While some cloud formation models suggest that a large-scale galactic magnetic field is irrelevant at the scale of molecular clouds (e.g., {\em Dobbs} 2008), because the turbulence and rotation of a cloud may randomize the orientation of its B-field (Figure 2, right panel), other models (e.g., {\em Shetty and Ostriker} 2006) have envisioned a galactic B-field strong enough to channel cloud accumulation and fragmentation (Figure 2, left panel). Recent observations have shed light on this debate.

\bigskip
\noindent
\textbf{ Observation I: Spiral-arm B-Fields Anchor Giant Molecular Clouds}
\bigskip

A measurement of the field direction in individual clouds and comparison to the spiral arms should determine which model is correct, but this is difficult to perform in the Milky Way because the arms cannot be observed due to the edge-on view of the Galactic disk. Furthermore, state-of-the-art instruments are not sufficiently efficient to probe the cloud B-fields from a face-on galaxy with the conventional cloud B-field tracer, polarization of dust thermal emission. Thus {\em Li and Henning} (2011) tried to probe cloud fields in M33 with the polarization of CO line emission (due to the Goldreich-Kylafis effect), which is much stronger than thermal dust emission. One problem with the Goldreich-Kylafis effect ({\em Goldreich and Kylafis} 1981; {\em Cortes et al.} 2005) is that a line polarization can be either perpendicular or parallel to the local B-field direction projected on the sky. Their argument that the 90-degree ambiguity can still be statistically useful is as follows. An intrinsically flat field distribution, as happens when the turbulence is super-Alfv\'enic or the cloud is rotating, will still be flat with this ambiguity. On the other hand, an intrinsically single-peaked Gaussian-like field distribution, as happens when the turbulence is sub-Alfv\'enic, will either stay single-peaked, or split into two peaks approximately 90-degrees apart.

M33 is the nearest face-on galaxy with pronounced optical spiral arms. The sub-compact configuration of the Submillimeter Array (SMA) offers a linear spatial resolution of $\sim 15$ parsecs at 230 GHz (the frequency of the CO J=2-1 transition) at the distance of M33 (900 kpc). {\em Li and Henning} (2011) picked the six most massive clouds from M33 for their strong CO line emission. The distribution of the offsets between the CO polarization and the local arm directions clearly shows double peaks (Figure 3). The distribution can be fitted by a double-Gaussian function with the two peaks lying at $-1.9^{\circ}\pm 4.7^{\circ}$ and $91.1^{\circ}\pm 3.7^{\circ}$ and a standard deviation of $20.7^{\circ}\pm 2.6^{\circ}$. This indicates that the mean field directions are well-defined and highly correlated with the spiral arms, which is consistent with the scenario that galactic B-fields can exert tension forces strong enough to resist cloud rotation (Figure 2, left panel).

\begin{figure}[ht]
\epsscale{0.5}
  \includegraphics[trim=2.2cm 0cm 2.1cm 0cm, clip=true, totalheight=0.28\textheight, angle=0]{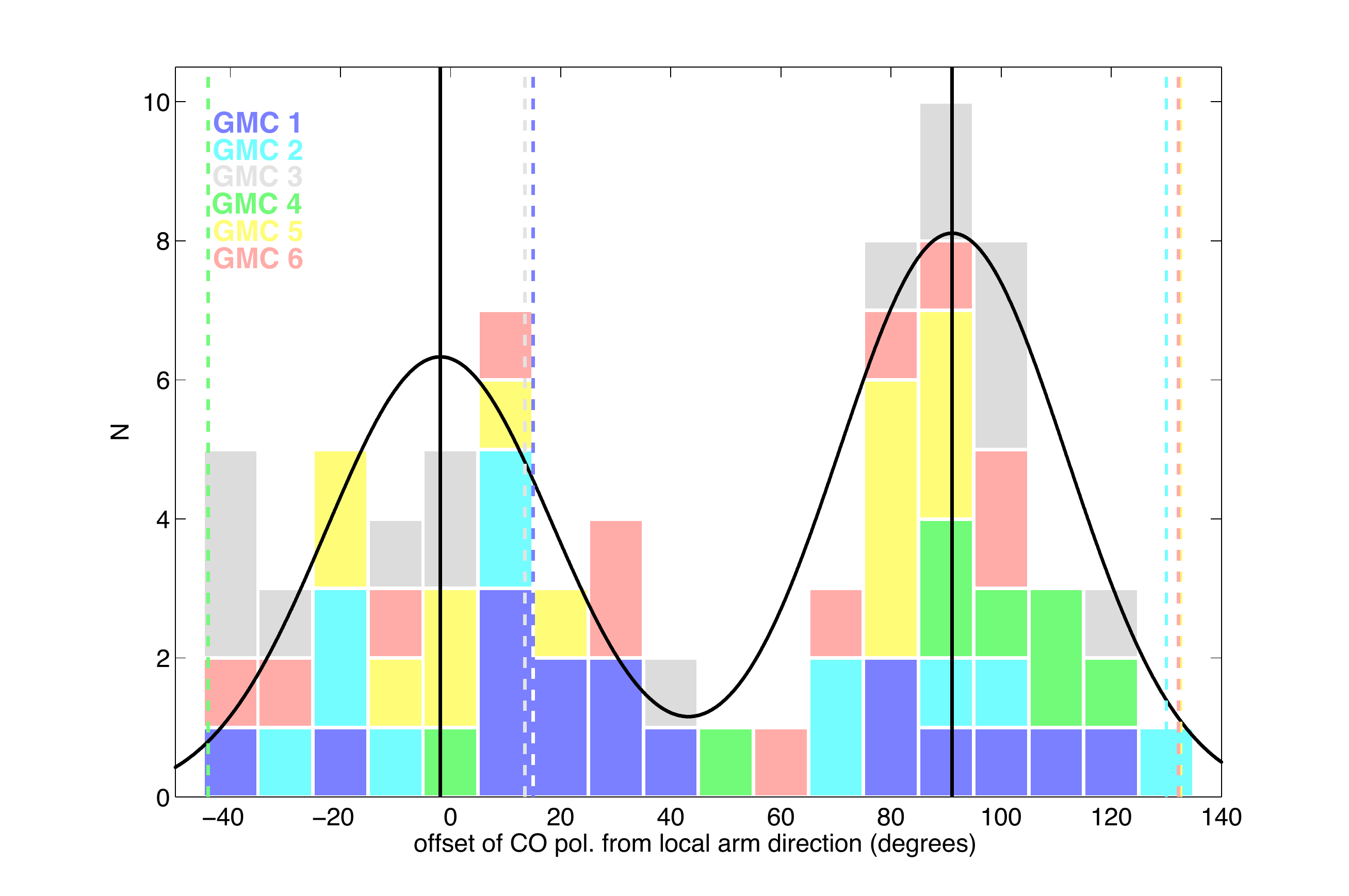}
 \caption{\small Distribution of the CO polarization-spiral arm offsets ({\em Li and Henning} 2011). The offsets are from the difference between the orientations of CO polarization and local arms in M33. Contributions from different GMCs are distinguished by the colors. The distribution can be fitted by a double-Gaussian function with a separation of $\sim 90^{\circ}$, suggesting that  the cloud B-field directions are correlated with the arm directions. The dashed lines show directions of synchrotron polarization, which are not relevant to the discussion here. For color figure see the arXiv version.
}  
 \end{figure}

\bigskip
\noindent
\textbf{ Observation II: Constant B-Field Strength from Inter-cloud Media to Clouds}
\bigskip

In a recent review of molecular cloud magnetic field measurements, {\em Crutcher} (2012) summarized the Zeeman measurements from the past decade in a B$_{los}$ (line-of-sight field strength)-versus-N$_{H}$ (H column density) plot. We show this plot in Figure 4 and highlight the column density ranges of inter cloud media (ICM) traced by HI data. Based on a Bayesian analysis, {\em Crutcher et al.} (2010) concluded that the two most probable scenarios for ICM B-field strength are: (1) constant strength around 10  $\mu$G and (2) any strength between 0 and 10 $\mu$G with a median of 6 $\mu$G. Most importantly, the B-field strength remains relatively constant in both cases over column densities from the ICM to the lower density regions in clouds. Independent analysis from {\em Heiles and Troland} (2005) on the same data shown in the ICM zone of Figure 4 also concluded a B-field strength median of $\sim 6\ \mu$G, independent of column density.
This is interpreted as evidence that gas can only accumulate along the B fields during cloud formation ({\em Crutcher et al.} 2010; {\em Crutcher} 2012), again consistent with Figure 2, left panel.

{\em Lazarian et al.} (2012) suggested an alternative scenario: the B fields were compressed during the cloud formation process, but the turbulence enhanced ''reconnection diffusion'' of B fields is efficient enough to keep the field strength independent of density. However, see Observation III.

 \bigskip
\noindent
\textbf{Observation III: Agreement between Cloud-contraction Threshold and Magnetic 
                              Critical Column Density}
\bigskip

Both Observation I \& II support the strong-B field scenario (left panel of Figure 2): cloud masses are first accumulated along B-fields. Only after a cloud has accumulated enough gas to reach the magnetic critical density can gravitational contraction happen in all directions. 
In Figure 4, for roughly N$_{H} > 10^{21} cm^{-2}$, the B-field strength increases with N$_{H}$, which implies that self-gravity is able to compress the field lines after accumulating adequate mass along the fields. The slanted solid line in Figure 4, $B\ (\mu G) = 3.8\times 10^{-21} N_{H}\ (cm^{-2})$ represents the so-called magnetic critical condition ({\em Crutcher} 2012; {\em Nakano and Nakamura} 1978): regions above the line can, in principle, be supported against their self-gravity by the magnetic forces alone. Since the cloud mass is accumulated along the B-fields, the cloud shape should be more sheet-like (e.g., {\em Shetty and Ostriker} 2006) instead of spherical, so statistically the observed column density should be twice the value to calculate the criticality (the column density observed with a sight line aligned with the B-field) due to projection effects ({\em Shu et al.} 1999). Taking the projection effects into account, we add to Figure 4 the corrected magnetic critical condition, 

$B\ (\mu G)= 1.9\times 10^{-21} N_{H, crit}\ (cm^{-2})$.\ \ \ \ \ \ (I)

Assuming an equipartition between turbulent and magnetic energies (upper limit of turbulent energy of sub-Alfv\'enic clouds), and magnetic virial equilibrium, 2T + M + U = 0 (where T, M and U are, respectively, kinetic, magnetic and gravitational potential energies; {\em McKee et al.} 1993), the critical condition becomes 

$B\ (\mu G)= 1.1\times 10^{-21} N_{H, crit}\ (cm^{-2}) $.\ \ \ \ \ \ (II)

Equations (I) and (II) give the lower and upper limits of the critical column density. For $B = 6-10\ \mu G$, the critical column density ranges between $N_{H} =[3.2, 9.1]\times 10^{21} cm^{-2}$ (Figure 4).

An indicator of gravitational contraction is the shape of a probability density function (PDF) of cloud column densities. Numerical simulations show that this PDF of non-gravitating clouds is log-normal for both sub- and super-Alfv\'enic clouds (e.g., {\em Li et al.} 2008a; {\em Collins et al.} 2012). When self-gravity is also accounted for, the PDF of high-density regions, where gravitational energy dominates other forms of energy, deviates from the log-normal function that describes the low-density PDF (e.g., {\em Nordlund and Padoan} 1999). This log-normal type PDF and the deviation are indeed observed ({\em Kainulainen et al.} 2009; {\em Froebrich and Rowles} 2010) and the transition point, cloud-contraction threshold ($N_{H}$  $= [3.7, 9.4]\times10^{21}$ cm$^{-2}$), is comparable with the critical column density ($N_{H}$  $= [3.2, 9.1]\times10^{21}$ cm$^{-2}$) ({\em Li et al.} 2013). We note that the empirical star-formation threshold, $N_{H}$  $= [1.0, 1.7]\times10^{22}$ cm$^{-2}$ ({\em Heiderman et al.} 2010; {\em Lada et al.} 2010), is a bit larger than the cloud-contraction threshold (Figure 4).

\begin{figure*}[float]
 \epsscale{1.75}
\plotone{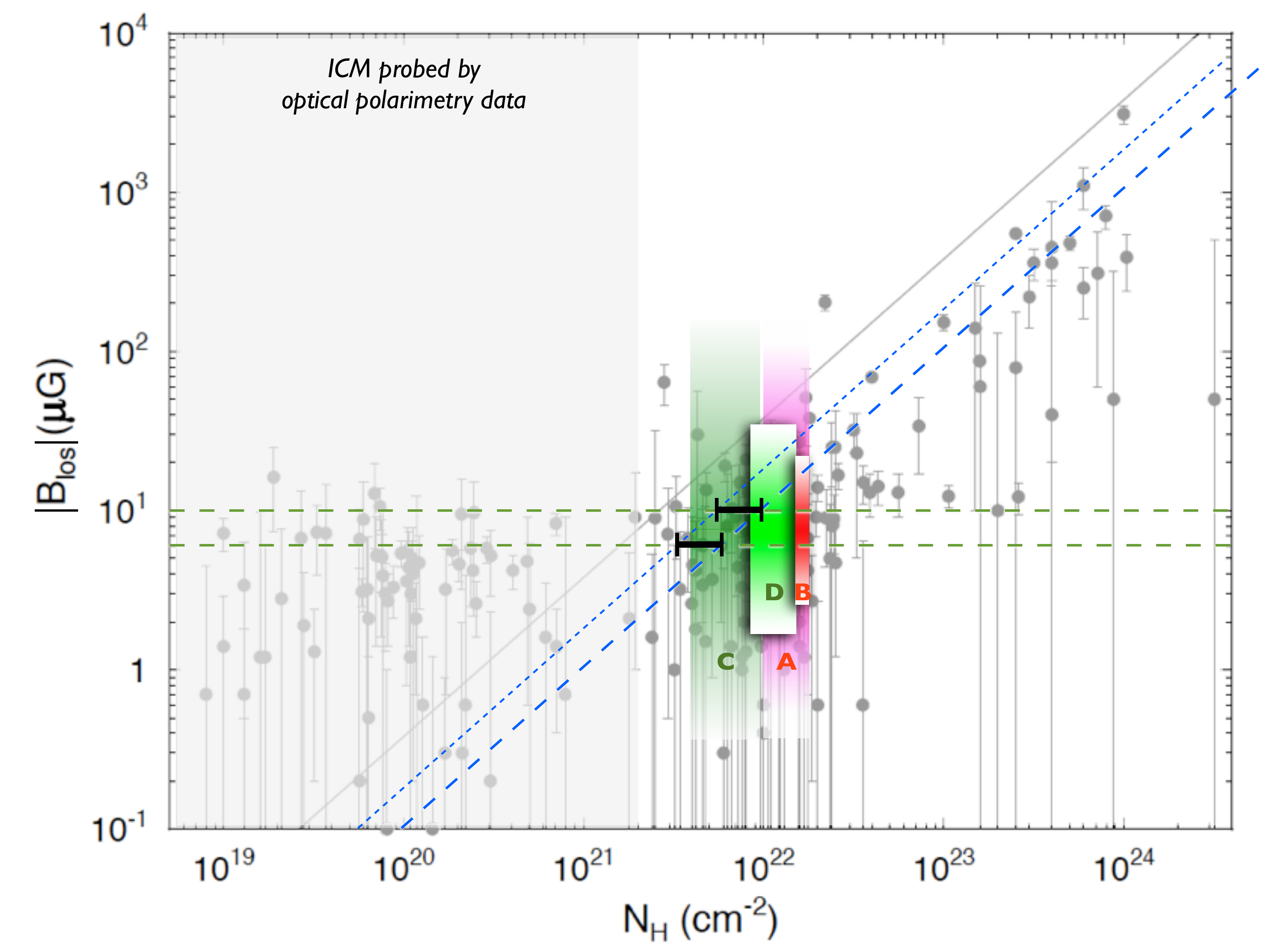}
 \caption{\small  The agreement between the magnetic critical density and gravitational contraction threshold. On top of the plot of line-of-sight field strength (B$_{los}$) versus column density (N$_{H}$) ({\em Crutcher} 2012), we indicate the inter molecular cloud media (ICM) by the light shaded zone, which is also the typical column density traced by optical polarimetry data (Observation IV \& V). The field strength is quite constant from the ICM to lower-density regions of the clouds ({\em Crutcher et al.} 2010). The two horizontal lines mark 10 $\mu$G and 6 $\mu$G, respectively ({\em Crutcher et al.} 2010).
The slanted solid line is the theoretical magnetic critical condition from {\em Crutcher} (2012). Applying a projection-effect correction ({\em Shu et al.} 1999) to it, we obtain the short-dashed line. Assuming an equipartition condition between magnetic and turbulent energies, we obtain the upper limit of the critical condition (slanted long-dashed line). The two ``H" shaped symbols mark the range of possible critical column densities for, respectively, B = 10 and 6 $\mu$G.
The emprical star formation threshold, A$_{V}$ = 7.3$\pm$1.8 mag from {\em Lada et al.} (2010) and  A$_{V}$= 8.1$\pm$0.9 mag from {\em Heiderman et al.} (2010) are shown by the zones labled A and B. Zone C ({\em Kainulainen et al.} 2009) and D ({\em Froebrich and Rowles} 2010) show where the observed column density PDF turns from log-normal to power-law like. For color figure see the arXiv version.     
}  
 \end{figure*}

 \bigskip

\centerline{\textbf{ 3. FILAMENT FORMATION }}
\bigskip
 
Recent observational studies, especially those exploiting the Herschel data ({\em Andr\'e et al.} 2010; {\em Molinari et al.} 2010; {\em Henning et al.} 2010; {\em Hill et al.} 2011; {\em Ragan et al. 2012}), have supported the decades old ``bead string" scenario of star formation ({\em Schneider and Elmegreen} 1979; {\em Mizuno et al.} 1995; {\em Nagahama et al.} 1998). In this picture, molecular clouds first form filaments (parsec- to tens-of-parsecs long ``strings"), which then further fragment into dense cores (``beads"). This scenario emphasizes the significance of filamentary structures as a critical step in star formation.

\begin{figure*}
\epsscale{1.6}[ht]
\plotone{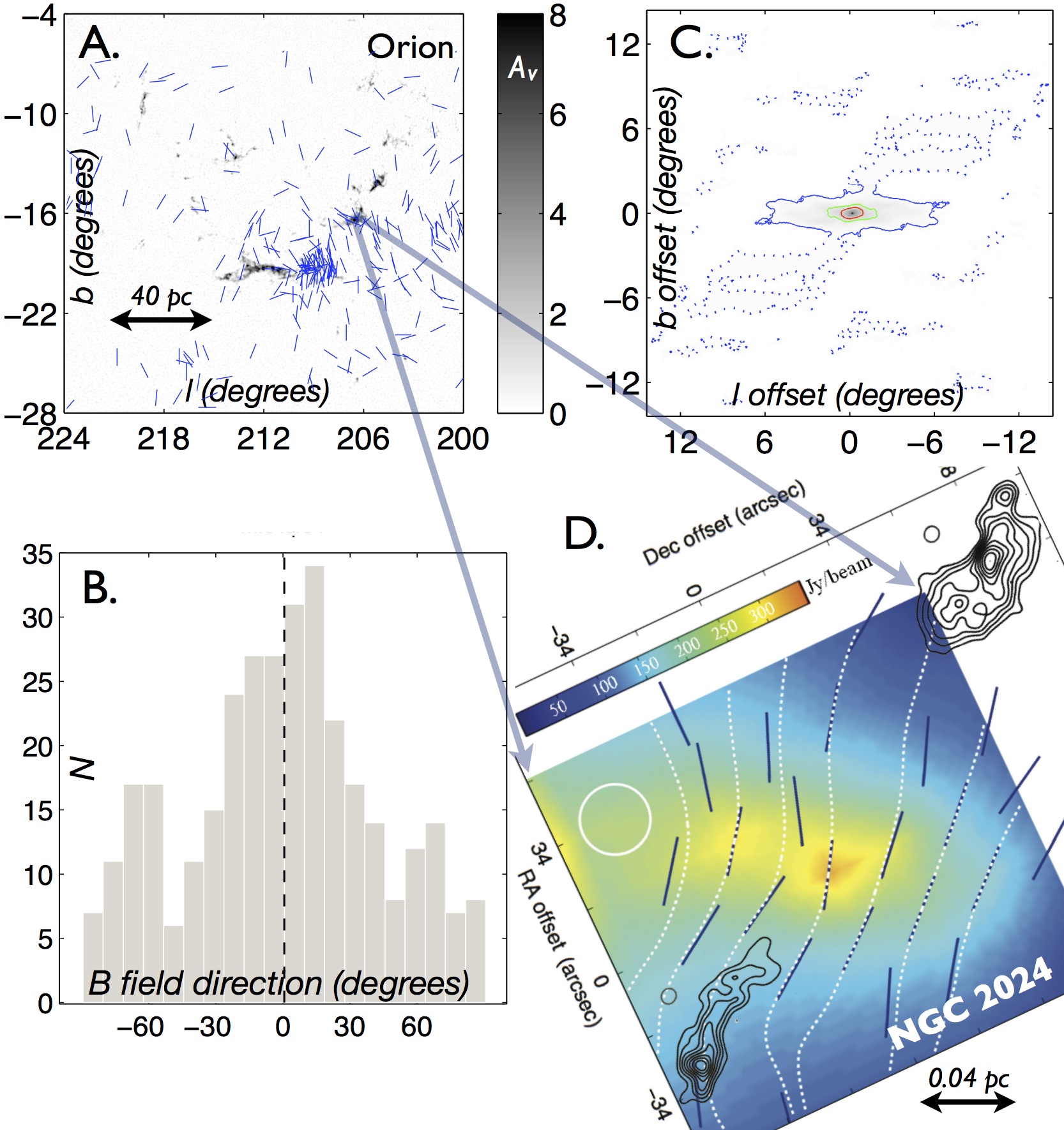}
\caption{
\small  (A) A$_{V}$ maps from {\em Dobashi} (2011) in Galactic coordinates (degrees) overlapped with optical polarization from {\em Heiles} (2000). 
\endgraf
\
 \endgraf
 (B) Distributions of the ICM B-field orientations inferred from the optical polarimetry data shown in panel A, measured counter-clockwise from Galactic North in degrees. The dashed line shows the Stokes mean of all the field detections in the map. 
 \endgraf
 \
 \endgraf
(C) The autocorrelation of the A$_{V}$ map shown in panel A. The coordinates are the offsets in degrees. The three contours are, respectively, of value 1 (blue), 1/5 peak value (red), and the mean of the previous two values (green).  The contour direction gives the cloud elongation. There are clearly several blue {\em parallel} contours and they are almost perpendicular to the mean field direction from (B). A survey of the correlation between cloud elongation (filaments) and mean field direction is given in Figure 8 (Observation V). 
 \endgraf
 \
 \endgraf
(D) Zoom-in on to the NGC 2024 core in panel A. The false-color map: dust thermal emission at 350$\mu$m ({\em Dotson et al.} 2010) with 20-arcsec resolution (white circle), centred on the FIR-3 core. Blue vectors: the magnetic field directions inferred from the polarimetry data. Dotted lines: magnetic field lines based on the polarimetry data ({\em Li et al.} 2010). 
Note that the field direction in panel A is very close to the field direction here, regardless of the very different scales and densities. A survey of the correlation between ICM and core B fields is given in Figure 6 (Observation IV). Also note that the B field is almost perpendicular to the core elongation; more discussion is given in Observation VIII.
Contours: HCO$^{+} (3-2)$ ({\em Li et al.} 2010). Note that the dense dust ridge is perpendicular to the mean field direction and that the HCO$^{+}$ streaks are parallel with the fields. Similar structures are seen at cloud (pc) scale and are called hub-filament structure (Observation VI; {\em Myers} 2009)
}
\end{figure*}
 
 \bigskip
\noindent
\textbf{ Models}
\bigskip

However, the formation mechanism behind filamentary clouds is still not understood. One popular model for filament formation is shock compression due to stellar feedback, supernovae, or turbulence (e.g., {\em Padoan et al.} 2001; {\em Hartmann et al.} 2001; {\em Arzoumanian et al.} 2011). However, this model is in contradiction with the fact that molecular clouds commonly show long filaments parallel with each other ({\em Myers} 2009). Stellar feedback (see, e.g., Figure 3 of {\em Hartmann et al.} 2001) and isotropic super-Alfv\'enic turbulence (see, e.g., Figure 2 of {\em Padoan et al.} 2001) cannot explain these parallel cloud filaments naturally. 
There are two other possible mechanisms to form filamentary clouds, which both require dynamically dominant B fields. These are B-field channeled gravitational contraction (e.g., {\em Nakamura and Li} 2008) and anisotropic sub-Alfv\'enic turbulence ({\em Stone et al.} 1998). In the former mechanism, the Lorentz force causes gas to contract significantly more in the direction along the field lines than perpendicular to the field lines, if the gas pressure is not strong enough to support the cloud against self-gravity along the B field ({\em Mouschovias} 1976). This contraction will result in flattened structures, which look elongated on the sky (e.g., {\em Nakamura and Li} 2008). When there are multiple contraction centers, the gas will end up in parallel filaments. 
Sub-Alfv\'enic anisotropic turbulence has the opposite effect: turbulent pressure extends the gas distribution more in the direction along the field lines, and leads to filaments aligned with the B field (see Figure 2 of {\em Stone et al.} 1998; {\em Li et al.} 2008a). This means that the competition between gravitational and turbulent pressures in a medium dominated by B fields will shape the cloud to be elongated either parallel or perpendicular to the B fields.

The presence of ordered dynamically important B fields required in the scenarios discussed above has recently received significant observational support.

  \bigskip
\noindent
\textbf{ Observation IV: Preservation of Cloud Field Directions}
\bigskip

It is hard to tell whether a cloud is globally super- or sub-Alfv\'enic by directly measuring the field strength. This is because both the conventional methods, i.e., Zeeman measurements (e.g., {\em Troland and Crutcher} 2008) and the Chandrasekhar-Fermi method (e.g., {\em Crutcher, Nutter and Ward-Thompson} 2004) measure only the strength of some field components, and have significant uncertainty in their estimates (e.g., {\em Hildebrand et al.} 2009; {\em Mouschovias and Tassis} 2009, 2010; {\em Crutcher, Hakobian and Troland} 2010; {\em Houde et al.} 2009).

A strategy to tackle this problem is to see whether cloud B fields are ordered (with well-defined mean direction) or random. It takes only a slightly super-Alfv\'enic turbulence to make a B-field morphology random in numerical simulations. The transition from ordered to random field morphologies is quite sensitive to the Alfv\'enic Mach number ($M_{A}$, the ratio of turbulent to Alfv\'enic velocity). For example, {\em Falceta-Gon\c{c}alves et al.} (2008) showed that B-field morphologies are ordered for $M_{A} = 0.7$ but random for $M_{A} = 2$ (no value in between was shown; see discussion in section 6.2). However, mapping the B field morphology of a {\em whole cloud} using the polarization of thermal dust emission was extremely time consuming. The SPARO team ({\em Novak} 2006) published the first four molecular cloud B-field maps with $\sim 5$ arcmin resolution ({\em Li et al.} 2006). Each cloud took about a month of telescope time from the Antarctic station. The field morphologies they observed are either shaped by the shells of H$_{II}$ bubbles or quite ordered.  

Other more efficient techniques had been developed to survey cloud fields. The plane-of-the-sky B-field orientations in inter cloud media (ICM) and in cloud cores can be measured using, respectively, optical and submillimetre polarimetry within a reasonable observation time (Figure 5) and polarization archives have been built for both cases. {\em Li et al.} (2009) compared the fields from 25 cloud cores (subpc scale), from the Hertz ({\em Dotson et al.} 2010) and SCUpol ({\em Curran and Chrysostomou} 2007) archives, to their surrounding ICM (hundreds of pc scale) fields direction inferred by the Heiles (2000) optical polarization archive and found a significant correlation (Figure 6): $90\%$ of the offsets between core and ICM B fields are less than $45^{\circ}$. The probability for obtaining this correlation from two random distributions is less than $0.01\%$.  This result agrees with {\em Li et al.}( 2006). Compared to the cloud simulations in the literature, only sub-Alfv\'enic ones present a similar picture.

A similar idea is to compare B$_{los}$ directions from cloud cores and from the ICM. Zeeman measurements of OH maser lines can probe core B$_{los}$, and Faraday rotation measurements of pulsars probe Galactic B$_{los}$. Comparing the two data sets, {\em Han and Zhang} (2007) concluded that B fields in the clouds still ``remember" the directions of Galactic ICM B-field directions.
In principle, Zeeman measurements of emission probe 3D field morphology for regions with $n(H_{2})  > 10^{5}   cm^{-3}$, but the interpretation of emission polarization is not straightforward ({\em Crutcher} 2012;  {\em Vlemmings et al.} 2011).

Nowadays, a survey of cloud B fields has been made possible by state-of-the-art instruments and their early results look consistent with SPARO and Figure 6. These surveys use either more efficient sub-mm polarimeters, e.g., BLAST-Pol ({Fissel et al.} 2010) and Planck ({\em Fauvet et al.} 2012), or use infrared polarimetry to fill the gap between cloud cores and ICM, e.g., MIMIR ({\em Clemens et al.} 2007) and SIRPOL ({\em Kandori et al.} 2006). Some examples are shown in Figure 7.

\begin{figure}[ht]
  \includegraphics[trim=0cm 0cm 0cm 0cm, clip=true, totalheight=0.4\textheight, angle=0]{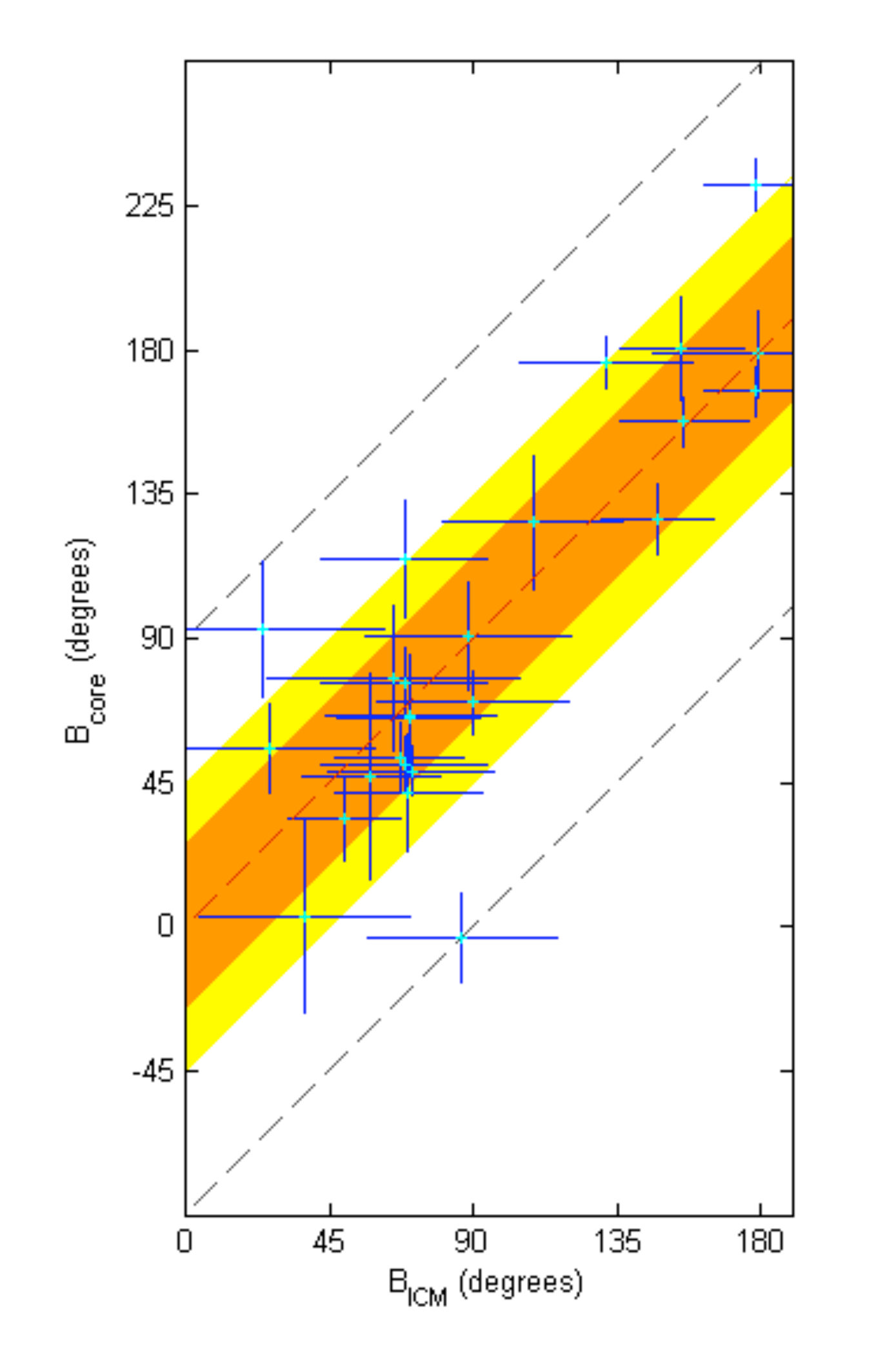}
 \caption{\small Correlation between B-field directions of cloud cores (B$_{core}$) and surrounding inter cloud media (B$_{ICM}$) ({\em Li et al.} 2009). The directions are measured from north$-$south in J2000 coordinates, increasing counterclockwise. The blue bars indicate the interquartile ranges (IQRs) of the B-field directions. The mean of the IQRs from all the ICM regions is approximately $52^{\circ}$ (orange area). About 70\% of the core/ICM pairs deviate from perfect parallelism by less than $26^{\circ}$. Nearly 90\% of the B$_{core}$ values are more nearly parallel than perpendicular to B$_{ICM}$ (orange and yellow
regions together).
}  
\end{figure}

\begin{figure*}[float]
  \includegraphics[trim=0cm 0cm 0cm 0cm, clip=true, totalheight=0.26 \textheight, angle=0]{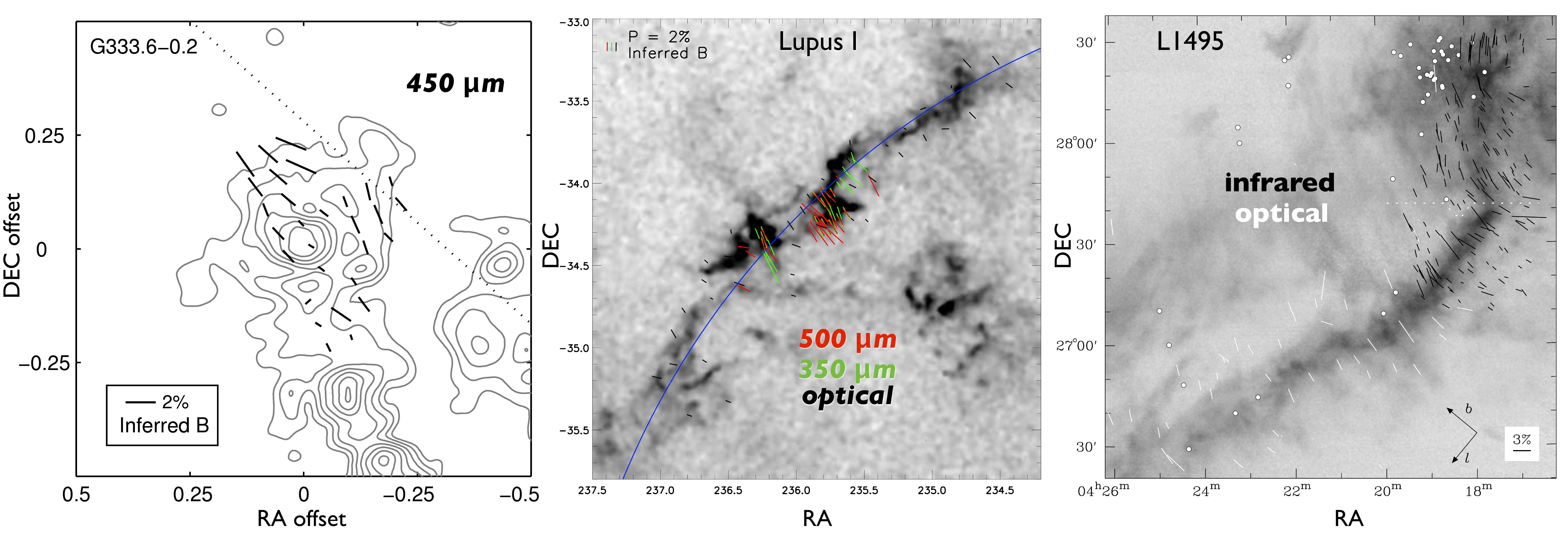}
 \caption{\small Examples of cloud B-field morphologies. Vectors with different colors show B-field directions probed by different wavelengths. {\em Left}: Sub-mm intensity contours overlapped by submm polarimetry inferred B-fields ({\em Li et al.} 2006). {\em Middle}: Inferred B-fields on top of the $A_{V}$ map. Optical and submm data are aligned ({\em Matthews et al.} 2013). {\em Right}: Greyscale shows relative $A_{V}$. Optical and infrared data are aligned ({\em Chapman et al.} 2011). For color figure see the arXiv version.
 }  
\end{figure*}

\bigskip
\noindent
\textbf{ Observation V: Bimodal Filament-Field Orientation}
\bigskip 

Observation I-IV tell us that molecular clouds are threaded by B-fields in well-defined directions, which means that cloud B-fields are not randomized by self-gravity or turbulence during the cloud and core formation processes. They are also not compressed in regions with $N_{H}$  $< 5\times10^{21}$ cm$^{-2}$, which comprise most of the volume of a molecular cloud ({\em Kainulainen et al.} 2009; {\em Lada et al.} 2010). This ordered B-field should, in turn, channel turbulent and gravitational gas motion, such that the resulting cloud shapes are elongated in directions either parallel or perpendicular to the local ICM B-fields. 
This theoretical model prediction was discussed in PPIII ({\em Heiles et al.} 1993), where, 
based on the study of Taurus, Ophiuchus and Perseus ({\em Goodman et al.} 1990), no significant alignment was found between the field orientations and cloud elongation. {\em Li et al.} (2013) revisited this issue using a larger sample, and came to a different conclusion. Their results are summarized below.
 
Cloud elongation and their local ICM B-fields can be probed respectively by extinction measurement (A$_{V}$; {\em Dobashi} 2011) and optical polarization data ({\em Heiles} 2000), as illustrated in Figure 5. {\em Li et al.} (2013) looked into these two archival data sets and compared cloud elongation and ICM B-fields from 13 regions along the Gould Belt. Their result is summarized in Figure 8: all pairs of mean fields and cloud directions fall within $30^{\circ}$ from being either parallel or perpendicular to each other. The probability for obtaining this correlation from two random distributions is less than $0.6\%$. Monte Carlo simulations and Bayesian analyses are performed to study the typical range of the 3-D offset from parallel and perpendicular alignments: the 95\% confidential range is 0-20 degrees. This indicates that ICM B-fields are strong enough to guide gravitational contraction to form flat condensations perpendicular to them (e.g., {\em Nakamura and Li} 2008) and strong enough to channel turbulent flows to result in filaments aligned with them (e.g., {\em Stone et al.} 1998; {\em Li et al.} 2008a). 

For cloud cores, recent SMA polarimetry survey (Zhang et al., in prep.) also observed a trend of bimodal alignment between core elongation and core B-fields. A different trend for cloud cores is also observed; see Observation VIII.

   \bigskip
\noindent
\textbf{ Observation VI: Hub-filament Cloud Structure}
\bigskip
 
Figure 8 covers all the nearby clouds examined by Myers (2009), who concluded that the clouds can be often described by a ``hub-filament" morphology: the clouds have high-density elongated ``hubs", which host most of the star formation in the clouds, and lower-density ``parallel filaments" directed mostly along the short axes of the hubs. Myers (2009) suggested that the parallel filaments are due to layer fragmentation. The main result of Observation V, i.e., that elongated/filamentary structures tend to be aligned either parallel or perpendicular to the ambient ICM B-fields, suggests that the gas layers in this scenario must host ordered and dynamically dominant B-fields, as shown by {\em Nagai et al.} (1998). 
Note that the two types of B-field-regulated filaments discussed in Observation V can also explain the hub-filament structures. The two types of filaments may form at the same time, with the denser and more massive filaments (hubs) perpendicular to the B-field and finer filaments in the vicinity aligned with the field (see observations from e.g., {\em Palmeirim et al.} 2013 and {\em Goldsmith et al.} 2008 and simulations from {\em Nakamura and Li} 2008 and {\em Price and Bate} 2008), which is a identical to the hub-filament structure described by {\em Myers} (2009). In this scenario, hubs and filaments should always be perpendicular to each other, but their sky projections are not necessarily so. This means that at least some of the exceptions, i.e., non-perpendicularity (e.g., Figure 9 of {\em Myers} 2009), can be explained by projection effects. The hub-filament system can be an explanation (besides rotation during contraction) of the larger differences between cloud directions defined by different column densities (e.g., Aquila in Figure 8). 

Hub-filament could be a self-similar structure. For example, Herschel Space Observatory resolved part of the Pipe nebula with 0.5-arcmin resolution and showed that a hub from {\em Myers} (2009) can fragment into a network of perpendicular filaments ({\em Peretto et al.} 2012). The network is aligned with the mean B-fields direction. An Observation V type analysis of the Herschel Gould Belt data will be of interest. {\em Li et al.} (2011) observed one of these core regions (NGC 2024; Figure 5) with 3-arcsec resolution and found filaments perpendicular to the core (aligned with the B-field), i.e., a hub-filament structure. A survey of core vicinities with high angular resolution and sensitivity as performed by {\em Li et al.} (2011) is necessary to tell whether NGC 2024 is a special case or not.

 \begin{figure}[ht]
  \includegraphics[trim=2.5cm 1cm 0cm 0cm, clip=true, totalheight=0.4\textheight, angle=0]{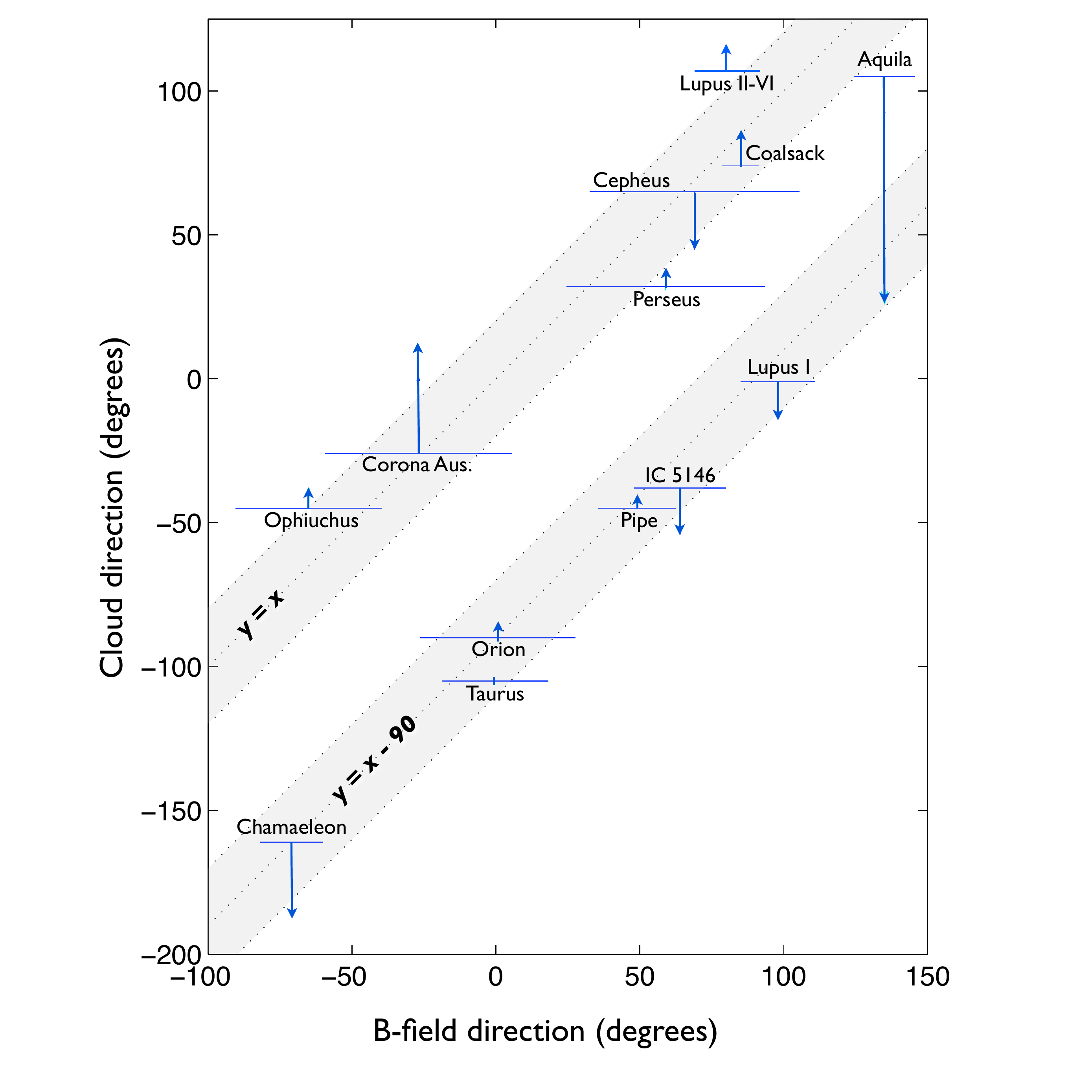}
 \caption{\small Cloud elongations versus B-field directions ({\em Li et al.} 2013). The range of directions of a cloud traced by the different column densities is shown by the vertical arrow, where the tail and head represent, respectively, the directions of the whole cloud and higher densities (1/5 peak value). The width ($42^{\circ}$) of the shaded regions in the X and Y directions is equal to the averaged value of the 13 IQRs (horizontal error bars) of the B-field directions. The filaments and B-fields tend to be either aligned or perpendicular to each other.
}  
 \end{figure}

    \bigskip
\noindent
\textbf{ Observation VII: Turbulent Anisotropy}
\bigskip

Given that cloud fields are ordered, turbulent velocities should be anisotropic. Turbulent energy cascades more easily in the direction perpendicular to the mean field than in the direction aligned with the field (i.e., along B-fields the velocity should be more coherent and less turbulent), if the field is dynamically important compared to the turbulence. This is because the development of turbulent eddies will be suppressed in the direction parallel to the field. This phenomenon should be more prominent at smaller scales, where turbulent energy is lower and field curvature (and thus tension) will be larger if the field is bent. An analytic model of anisotropic, incompressible magnetohydrodynamic (MHD) turbulence was first proposed by {\em Goldreich and Sridhar} (1995, 1997). {\em Cho and Vishniac} (2000), {\em Maron and Goldreich} (2001) and {\em Cho et al.} (2002) numerically verified the anisotropy predicted by the Goldreich-Sridhar model. Similar anisotropy was also observed in compressible MHD simulations from {\em Cho and Lazarian} (2002) and {\em Vestuto et al.} (2003).

{\em Heyer et al.} (2008) first observed turbulence anisotropy in a molecular cloud by showing that the $^{12}$CO $(1-0)$ turbulence velocity is more coherent in the direction along the B-field (Figure 9). The region they studied is roughly ($2^{\circ} \times 2^{\circ}$) centered at 4h50m0s, $27^{\circ}0'0''$ (J2000) in the Taurus Molecular Cloud. A$_{V}$ is less than 2 mag in this region ({\em Schlegel et al.} 1998), and thus the conclusion from this work is restricted to the cloud envelope. In the same region, they also observed that the $^{12}$CO emission exhibits ``filaments" (see observation V) that are aligned along the magnetic field direction. B-field channeled turbulent anisotropy is also seen with $^{12}$CO $(1-0)$ spectra from Ophiuchus ({\em Ji and Li in prep.}), where the direction with the least velocity dispersion tends to align with the B field.

{\em Heyer and Brunt} (2012) extended the study of Heyer et al. (2008) to higher density using $^{13}$CO $(1-0)$ lines, which traces  A$_{V} = 4-10$ mag and exhibits little evidence for anisotropy. {\em Heyer and Brunt} (2012) interpret the observation as that the turbulence transits from sub- to super-Alfv\'enic in regions with higher column density. However, we note that A$_{V} = 4$ mag is measured from the ``hub" (see observation VI) and happens to be the cloud-contraction threshold of TMC ({\em Kainulainen et al.} 2009; Observation III). It means that, over this density limit, contraction velocities in all directions dominate the turbulent velocity, so turbulent anisotropy is less clearly observable. Moreover, super-Alfv\'enic turbulence in regions with A$_{V} = 4-10$ mag also contradicts Observation IV: B-fields in densities below and above this density range are aligned. {\em Li et al.} (2011) observed the B-field aligned velocity anisotropy with $^{12}$CO $(7-6)$ emission from a region of NGC 2024 with the mean  A$_{V} = 7$ mag, after carefully examining the line profiles away from regions with contraction/expansion signatures. If the angle dispersion of a polarimetry map stems from turbulence, the angle dispersion should also be anisotropic if the turbulence is. This is exactly what {\em Chitsazzadeh et al.} (2011) found for OMC-1, which has A$_{V}$ above 30 mag ({\em Scandariato et al.} 2011). Given the available data, it is hard to tell whether there is a density limit for turbulent anisotropy.

 \begin{figure}[ht]
  \includegraphics[trim=0cm 0cm 0cm 0cm, clip=true, totalheight=0.34\textheight, angle=0]{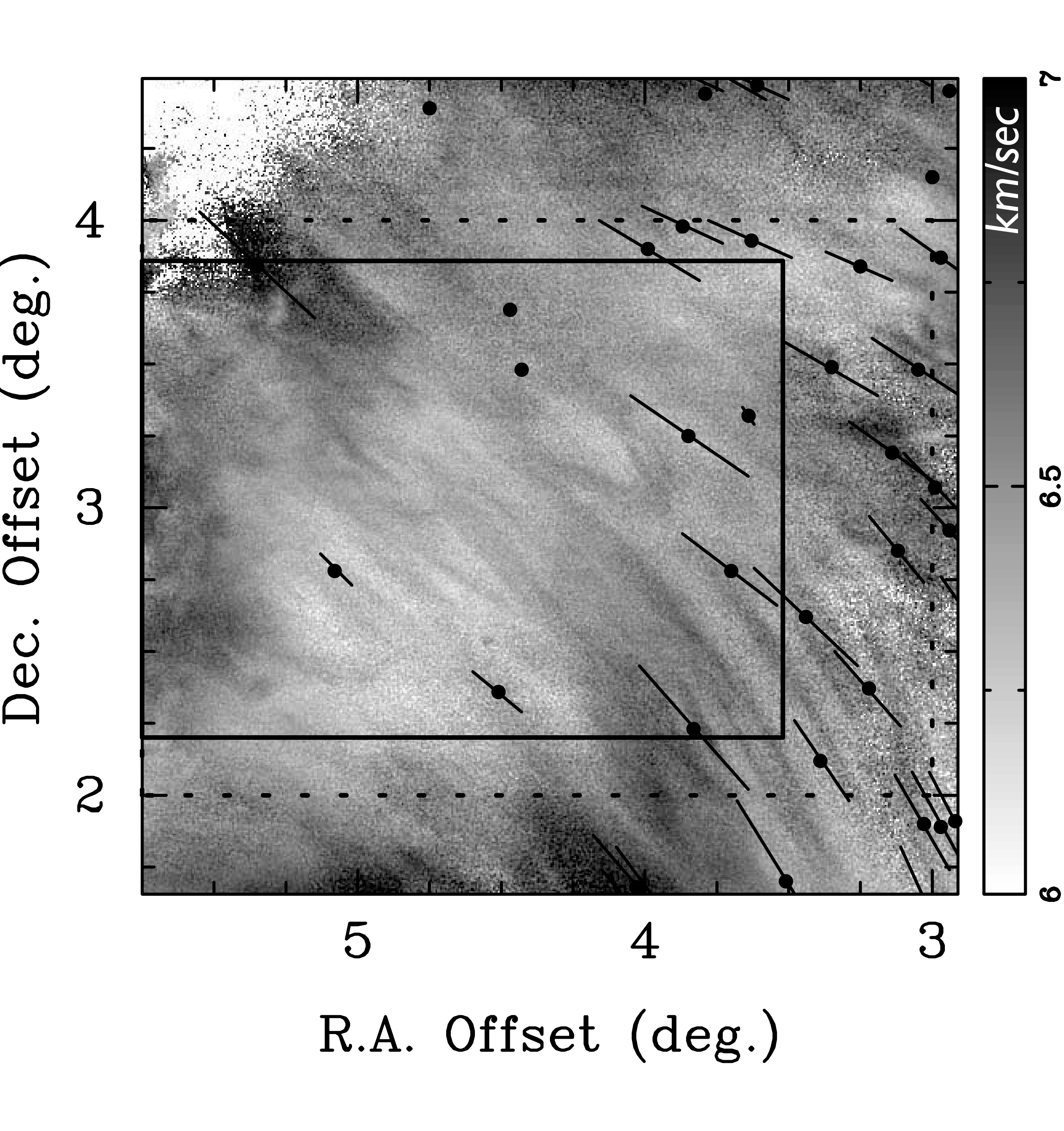}
 \caption{\small Image from {\em Heyer et al.} (2008) showing $^{12}$CO velocity centroid overlapped with B-field directions inferred from the optical polarimetry data from {\em Heiles} (2000). The velocity streaks are aligned with the B-field direction. The solid line box and dotted-line box show the areas within which the velocity anisotropy and B-field direction are estimated in  {\em Heyer et al.} (2008).  
}  
 \end{figure}
 
 \bigskip

\centerline{\textbf{ 4. CLOUD CORE FORMATION }}
\bigskip
  By cloud cores, we mean structures like in Figure 5, panel D, typically with linear scales of sub-pc and mean density $n_{H}$ $= 10^{5-6}$ cm$^{-3}$. They are nurseries of star formation.
  
    \bigskip
\noindent
\textbf{ Observation VIII: Cloud Cores Shaped by B Fields}
\bigskip  
  
Given that cloud fields and ICM fields are aligned, flux freezing should make a gravitational contraction anisotropic, and (somewhat) flattened high-density regions perpendicular to the mean field direction should be expected, as observed in the NGC 2024 core region (Figure 5). Panel D of Figure 5 covers the same region as {\em Crutcher} (1999) analysed for NGC2024. He estimated that the mass-to-flux ratio here is highly supercritical (4.6 times of the critical value), the highest among the 15 core regions (N$_{H} > 10^{23}$) with reliable detections in Figure 4. Even so, the magnetic field still clearly channels the contraction to form an elongated core with the long axis perpendicular to the field (Figure 5).

Observation of this kind of correlation, however, is not always
straightforward, because the real core flatness is not always
observable due to the projection effect. Observations suggest that
core projections are not generally spherical (e.g., {\em Benson
and Myers} 1989) and that the most probable core shapes are nearly
oblate (e.g., {\em Jones, Basu and Dubinski} 2001). If the shortest axis of an
oblate core indeed orients close to the mean field direction, then
the closer the line of sight is to perpendicular to the shortest
axis, the better the alignment should be observed between
the field projection and the short axis of the core projection. In
other words, the more elongated the core projection, the better the
alignment should be. This is exactly the trend {\em Tassis et al.} (2009)
observed from 32 cloud cores (Figure 10) with the CSO. 

\bigskip

 \begin{figure}[ht]
  \includegraphics[trim=0cm 0cm 0cm 0cm, clip=true, totalheight=0.25\textheight, angle=0]{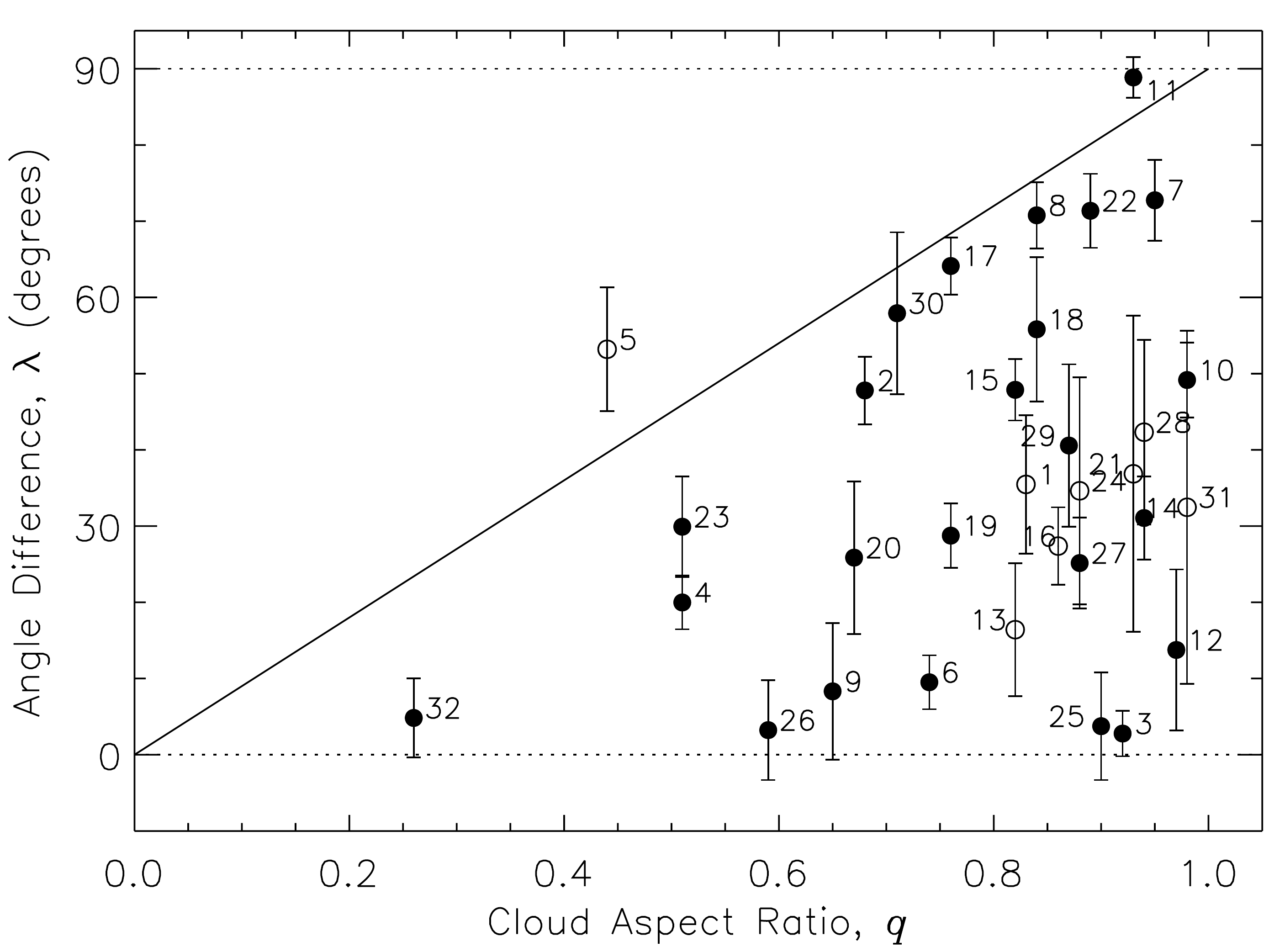}
 \caption{\small The alignment between the B-fields and the short axes of cores ({\em Tassis et al.} 2009). $\lambda$  is the angle difference between the sky projection of the {\em core B-field} and the {\em short axis of the core projection}. $q$ is defined as the short-to-long axis ratio of a core projection. The model suggested by {\em Tassis et al.} (2009) is that cores form under dynamically important B fields will be flattened such that the core short axes will align with the B-fields in 3D. This geometry is most easily observed (small $\lambda$) when the line of sight is close to perpendicular to the B-fields, or, in other words, when the flattened core is viewed edge-on (when $q$ is small). The upper limit of  $\lambda$ indeed increases with $q$, i.e., most data points are below the solid line ($\lambda/90^{\circ} = q$). The probability to get only 2 out of 32 data points above the solid line by chance is only $\sim 10^{-7}$.   
} 
\end{figure}

    \bigskip
\noindent
\textbf{ Observation IX: No Detection of Column Density-to-B$_{los}$ Ratio Increasing from Cloud to Cores}
\bigskip 

The proposal that B-fields can regulate star formation through ambipolar diffusion (AD) has been put forth for more than five decades ({\em Mestel and Spitzer} 1956; {\em Shu et al.} 1999; {\em Ciolek and Mouschovias} 1993), but observers had been unable to even try to test it untill the pioneer work of {\em Crutcher et al.} (2009).  If ambipolar diffusion plays a major role in core formation, increasing of mass-to-flux ratio from clouds to cores should be expected. With OH zeeman measurements, {\em Crutcher et al.} (2009) used {\em column density-to-B$_{los}$ ratios} to estimate mass-to-flux ratios. Comparing the mass-to-flux ratios between cores and surrounding envelopes from four regions (B217-2, L1544, B1 and L1448), they reported that the ratios of cores are smaller than the envelopes for all cases. They thus concluded that the result is more in agreement with the super-Alfv\'enic cloud simulation from {\em Lunttila et al.} (2008), which is without ambipolar diffusion and can produce cores with smaller  column density-to-B$_{los}$ ratios compared to their envelopes.

However, other than AD, there are uncontrolled factors in this experiment that also affect column density-to-B$_{los}$ ratios. These factors make it difficult for this experiment to discriminate or support any core formation mechanism. In the following we discuss these factors:

\bigskip

(1) {\em B$_{los}$  reversal}

Starless cores are usually more quiescent than their envelopes ({\em  Benson and Myers} 1989; {\em Goodman et al.}1998; {\em Barranco and Goodman} 1998). The stronger the turbulence, the larger the B-field dispersion ({\em Chandrasekhar and Fermi} 1953), and the more field reversal might happen within a telescope beam to {\em reduce} the mean B$_{los}$ measured by Zeeman effect. The stronger turbulence in envelopes can bias the column density-to-B$_{los}$ ratios toward the higher end.  For sub-Alfv\'enic turbulence, B-field direction has a Gaussian-like dispersion and the field reversal within a beam depends on the angle between the line of sight and the mean field direction. For example,  a gaussian B-field distribution with a STD = $30^{\circ}$ and a mean direction $60^{\circ}$ from the line of sight will have $\sim 17\%$ of the B$_{los}$ being flipped. This observational bias is illustrated by the sub-Alfv\'enic simulation shown in Figure 11 ({\em Bertram et al.} 2012), where the column density-to-B$_{los}$ ratios from the cores and from the envelopes are compared along various lines of sight. 

By definition, mass-to-flux ratios should be estimated by column density-to-B$_{los}$ ratios measured along the B-field direction (Z axis in Figure 11). But {\em Crutcher et al.} (2009) proposed that the ratio ($R$) between the column density-to-B$_{los}$ ratios from a core and from its envelope should be independent of lines of sight, assuming that B-fields from the core and from the envelope are aligned. Based on Observation IV, this seems to be a fair assumption. However, it ignores the fact that larger misalignment between B-fields and the line of sight produces more B$_{los}$ flips to disperse the measurements of B$_{los}$. With a line of sight aligned with the mean field (no B$_{los}$  reversal), Figure 11 shows that all cores have mass-to-flux ratios {\em larger} than their envelopes.  Along other lines of sight, field reversal can significantly lower the B$_{los}$ of many envelopes and give $R < 1$. Figure 11 shows that sub-Alfv\'enic turbulence is enough to flip B$_{los}$; super-Alfv\'enic turbulence ({\em Lunttila et al.} 2008) is not a necessary condition for observing $R < 1$. {\em Bertram et al.\ }(2012) also performed super-Alfv\'enic simulations and obtained results similar to Figure 11, only with more $R$ less than 1. But in super-Alfv\'enic cases the distribution of green data is not too different from red and blue data ({\em Bertram et al.} 2012) due to tangled B-field lines, which disagrees with Observation IV. Sub-Alfv\'enic turbulence introduces structures to B-fields without changing the mean direction, which can produce the R scattering in Figure 11 as well as the multi-scale field alignment shown in Figure 6.
\\ 
 
  \begin{figure*}[float]
\epsscale{1.5}
\plotone{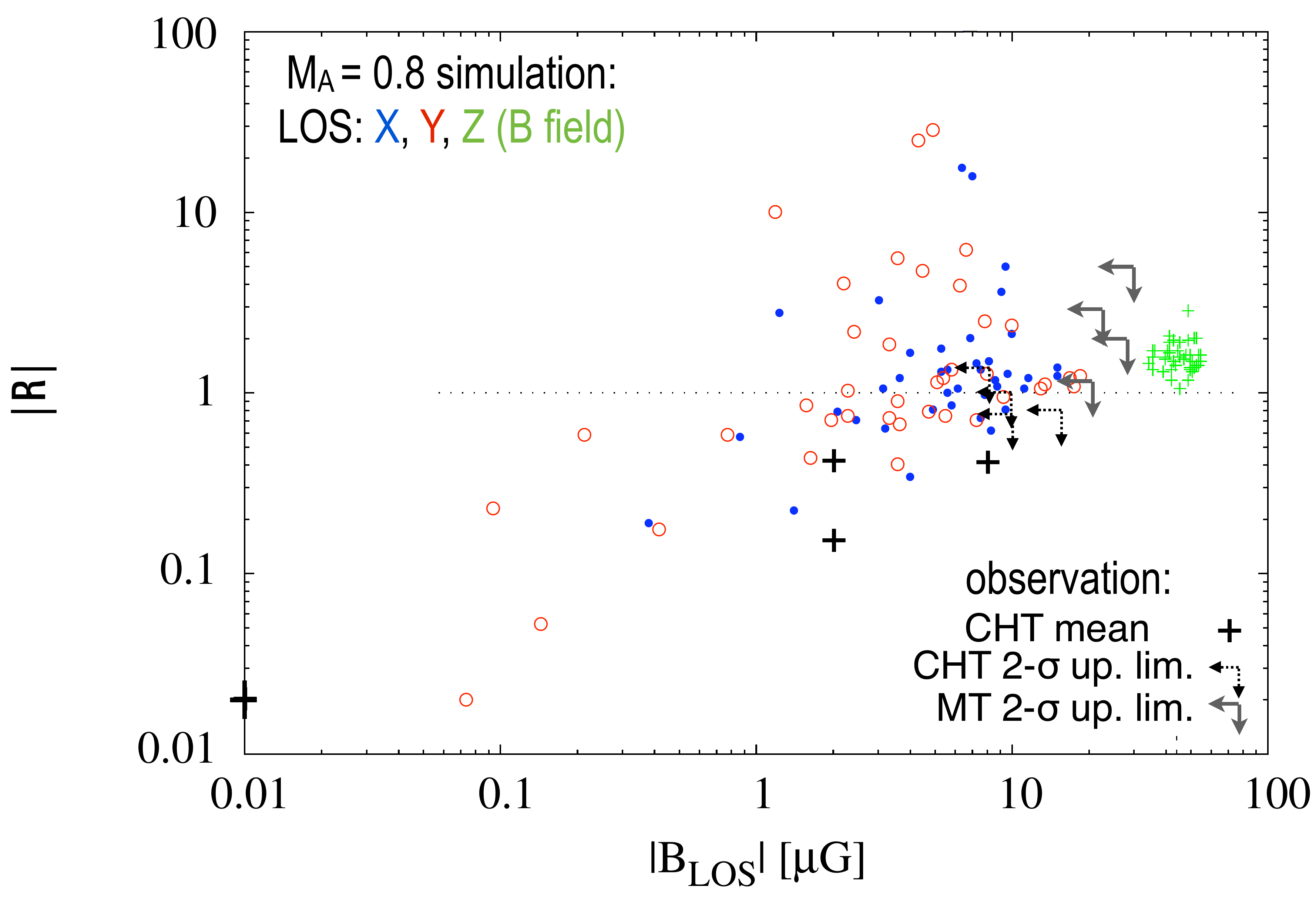}
 \caption{\small  Distribution of clumps from the sub-Alfv\'enic simulation ({\em Bertram et al.} 2012) and from observations ({\em Crutcher et al.} 2009; CHT). The simulated data is projected along different line-of-sight directions (X, Y, and Z axes) to compare with observations. Plotted is the absolute value of R, the ratio between the column density-to-B$_{los}$ ratios from the cores and those from the envelopes, against the absolute value of the average of B$_{los}$. {\em Only the column density-to-B$_{los}$ ratios measured along the mean B field (Z axis; green data points) are equivalent to the mass-to-flux ratios by definition}.  The initial magnetic field strength is almost unchanged when observed along the B-field, meaning that the B-field is {\em not tangled} and turbulence only introduces field dispersion. This field dispersion can be observed as B$_{los}$ reversal along other lines of sight and produces a large dispersion of R compared to those observed along the B-field. Observations from {\em Crutcher et al.} (2009) are also plotted, with the 2-sigma upper limits from  {\em Crutcher et al.} (2009) and from {\em Mouschovias and Tassis} (2009; MT). The latter authors analyzed the same set of observational data with different assumptions (see text). For color figure see the arXiv version.
}  
 \end{figure*}

(2) {\em Anisotropic gravitational contraction}

Figure 11 also shows that $R > 1$ is possible even {\em without} AD.  While mass is accumulated more along the B-field lines due to anisotropic contraction (Observation VIII) or anisotropic turbulent shocks (Observation VII) channeled by dynamically important B-fields, one should observe larger column density-to-B$_{los}$ ratios of cores ($R > 1$) even without ambipolar diffusion. This is because the column density increases faster than field strength does in cores due to the anisotropic gas contraction/compression mainly along the field lines ({\em Li et al.} 2011).  Only a compression perpendicular to the field lines can increase the B-field strength.
Like factor (1), the larger the offset between the line of sight and the B-field, the stronger factor (2). 
\\

(3) {\em Reconnection diffusion}

{\em Lazarian et al.} (2012) proposed that turbulent eddies can bring B-field lines to cross each other and get reconnected, and this reconnection can cause field diffusion with an efficiency increasing as turbulent energy increases. If cores are more quiescent than their envelopes, the latter will have a higher reconnection diffusion efficiency and thus higher mass-to-flux ratio ($R < 1$).
\\

While AD and factor (2) tend to make $R > 1$, factors (1) and (3) tend to make $R < 1$. The competition between these factors can produce a spectrum of $R$. This is seen in Figure 11 already (blue and red data points) even without AD. Note that only the green data points in Figure 11 are the ratios between the mass-to-flux ratios from  cores and from envelopes, the quantities {\em Crutcher et al.} (2009) planned to measure. In reality, however, the column density-to-B$_{los}$ ratios, which are the values {\em Crutcher et al.} (2009) used to estimate mass-to-flux ratios, will give biased $R$s, like the blue or red data points in Figure 11. Bearing in mind that B$_{los}$ may reverse from position to position to cause the scattering in the Zeeman measurements, {\em Mouschovias and Tassis} (2009) reanalyzed the data from {\em Crutcher et al.} (2009) (though their reason for reversal is not turbulence). They obtained a much higher possibility for $R > 1$ compared to the analysis from {\em Crutcher et al.} (2009), who assumed that the variation in their Zeeman measurements are mainly due to observation errors ({\em Crutcher et al.} 2010). 

An interesting fact about the simulations 
from {\em Bertram et al.} (2012) is that the mean of all $R$s is always larger than one, regardless of the magnetic Mach number. This is different from the results of the 4 observations from {\em Crutcher et al.} (2009). Whether this discrepancy is physical or due to the small observation sample size remains a question. On the other hand, super-Alfv\'enic simulations from {\em Lunttila et al.} (2008) indeed produce $\sim 80\%$ of the clumps with $R < 1$. However, their $|R|$ tends to be inversely proportional to B$_{los}$, which is again different from the trend of the 4 observations (Figure 11) and different from the super-/sub-Alfv\'enic simulations from {\em Bertram et al.} (2012). While more work is needed to better understand simulations and observations, one thing is obvious: with or without AD, $|R|$ can be either larger or less than unity.

 \bigskip

\centerline{\textbf{ 5. PROTOSTELLAR DISC FORMATION }}
\bigskip
The competition between angular momentum and magnetic braking, as shown in Figure 2, should  also take place at the scale of protostellar disc formation. Compared to the scale of cloud formation (Figure 2), here the angular momentum is weaker and magnetic braking is stronger. Galactic shear can be ignored at this scale and turbulence is also much weaker based on Larson's Law, and B fields are stronger based on Figure 4. Yet, unlike Observation I, here we know that angular momentum wins the competition in many cases where discs are observed (e.g., {\em Murillo and Lai} 2013; {Tobin et al.} 2012), so how it can win becomes a mystery. 

Due to the lack of instrumental sensitivity, observations of B fields
on scales of starless cores to discs is far behind the development of related theories.
ALMA should offer adequate resolution and sensitivity to probe the B fields of
protostellar disks and the dense cores they are embedded in. Observational constraints on
competing numerical models are extremely important, because the models of disk formation are not converging. In some models, efficient B-field
diffusion (e.g., {\em Li, Krasnopolsky and Shang} 2011) is indispensable for solving the, so-called, magnetic braking catastrophe, i.e., the B field tension force will prevent disk
formation if field lines are well coupled to the dense cores. Other groups, for
example, claim that turbulence (e.g., {\em Seifried et al.} 2012) or mass rearrangement from
the core to the pseudo-disc ({\em Machida, Inutsuka and Matsumoto} 2011) can mitigate the
effect of magnetic braking and thus enable disc formation. Numerical (artificial) effects
have not been ruled out as a cause for the differences between these various models. The chapter by Z.-Y. Li et al. in this volume has a detailed review of the disc formation theories. Here we review recent observations focusing on the role played by B fields in disc formation.
 
\bigskip
\begin{figure*}
 \epsscale{1.6}

\plotone{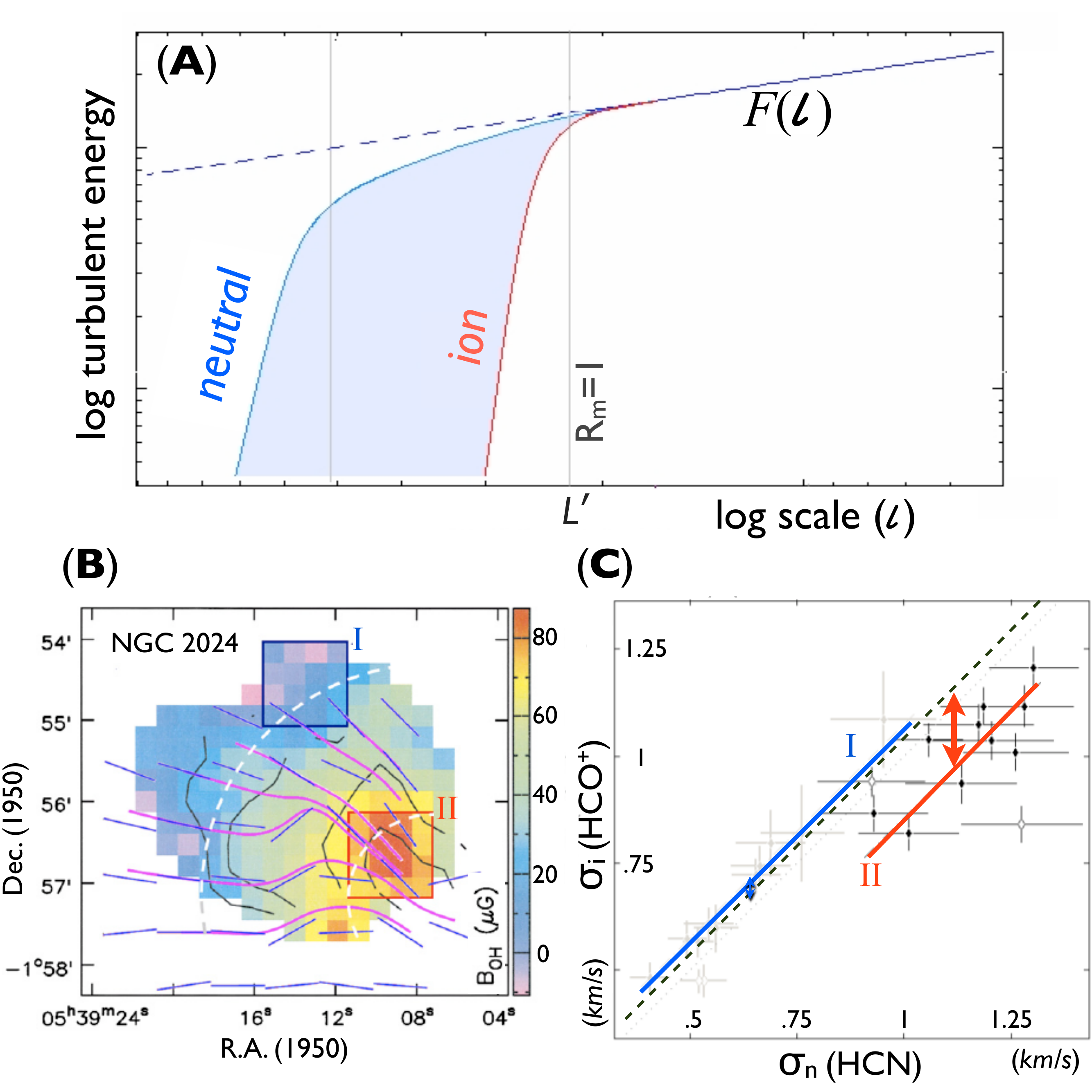}
 \caption{\small  (A) An Illustration of the difference between the ion and neutral velocity spectra, from {\em Li and Houde} (2008). This is a log-log plot of the turbulent energy
versus scale size in arbitrary units. The decoupling between neutrals and B fields
happens at scales smaller than $L^{\prime}$, where $R_{M} < 1$. The energy spectrum in the inertial range is common for both species, while ions and neutrals have different spectra at scales smaller than $L^{\prime}$. The square of the velocity dispersion ($\sigma$) measured at a particular scale (beam size) is proportional to the integral of the energy spectrum over all scales smaller
than the beam size. At scales larger than $L^{\prime}$, the observed difference between the two velocity dispersion spectra is proportional to the shaded area. 
 \endgraf
 \
 \endgraf
(B) An example of the correlation between ion/neutral velocity dispersion difference and B-field
strength. The colors show B$_{los}$ in NGC 2024, estimated by OH Zeeman
measurements ({\em Crutcher et al.} 1999). The vectors indicate the magnetic field directions
inferred from 100 $\mu$m polarimetry data ({\em Crutcher et al.} 1999). The solid curved lines are the magnetic field lines ({\em Li et al.} 2010). Velocity dispersions from the boxed regions (I and II) are shown in (C). 
 \endgraf
 \
 \endgraf
(C) HCO$^{+}$ (3-2) vs. HCN (3-2) velocity dispersions ($\sigma_{i}$ and $\sigma_{n}$). The maps of HCO$^{+}$ (3-2) from the boxed regions (Zone I and II) are shown in panel D of Figure 5. The light and dark colors are for, respectively, weak field (zone I) and strong field (zone II) in (B). The dashed line is X = Y. Note that the region with the stronger field has larger difference, on average (shown by the double-headed arrow) between $\sigma_{i}$ and $\sigma_{n}$ ({\em Li et al.} 2010).
}  
 \end{figure*}
 
 \bigskip

\noindent
\textbf{ Observation X: Decoupling between B Fields and Turbulence, a New Model of Ion-neutral Line Width Difference}
\bigskip 

For the scales of galactic disks and molecular clouds, flux freezing (the perfect coupling between gas and
magnetic field lines) is a good approximation, based on the MHD induction equation. Indeed, this is
why weak fields are expected to follow cloud rotation and turbulence, whilst strong fields will
channel turbulence and suppress cloud rotation (Figure 2). Moving toward small scales, however,
the MHD induction equation also suggests decoupling between turbulent eddies and B fields (when the magnetic Reynolds number, $R_{M}$, is approximately below unity; {\em Li and Houde} 2008). This may help to solve the magnetic braking catastrophe, if this decoupling scale, $L^{\prime}$, is comparable to the scales of protostellar discs (a few hundred AU) and turbulence at this scale is the main energy source of disc angular momentum. Below the decoupling scale, ions and neutrals should have different turbulent velocity (energy) spectra (Figure 12), because ions are still coupled with field lines due to the Lorentz force. 
To identify $L^{\prime}$ is challenging because it could be on the order of milli-pc ({\em Li and Houde} 2008), which is near the resolution limit of current radio and sub-mm telescopes. However, since all the turbulent eddies within a telescope beam contribute to the observed line width (velocity dispersion), the decoupling should still affect the line width even if their beam size is larger than $L^{\prime}$. In other words, the line widths of coexistent ions and neutrals should be different due to neutral gas-field decoupling. Indeed, many other factors, e.g., opacities, spatial distributions, hyperfine structures and outflows, can affect line widths, but these non-magnetic factors can be ruled out by carefully choosing the regions of observation and the ion/neutral pairs. HCO$^{+}$/HCN and H$^{13}$CO$^{+}$/H$^{13}$CN
can be used for this purpose ({\em Li and Houde} 2008; {\em Hezareh et al.} 2010), because the ion species generally have larger opacity, slightly more extended distribution and more unresolved hyperfine structures ({\em Li et al.} 2010); all of these factors suggest a wider line width, but ion spectra are systematically narrower ({\em Houde et al.} 2000 a \& b 2002). Most importantly, {\em Li et al.} (2010) have shown that the line-width difference between HCO$^{+}$ and HCN is proportional to the B field strength (Figure 12). This definitely cannot be explained by any reason other than {\em decoupling between B fields and small turbulent eddies}. The decoupling is called turbulent ambipolar diffusion (TAD) in {\em Li and Houde} (2008) to distinguish it from AD, the decoupling between B fields and gravitationally contracting gas (Observation IX).

By setting $R_{M}$ (the ratio between the advection term and diffusion terms of the induction equation) equal to unity, it can be shown that ({\em Li and Houde} 2008):

$B_{pos}^{2} \approx 4\pi n_{i}\mu v_{i}V_{n}^{\prime}L^{\prime} $,\\
where $B_{pos}$, $n_{i}$, $v_{i}$ and $\mu$ are, respectively, the plane-of-sky B-field strength, the ion density, the collision rate between an ion and the neutrals, and the reduced mass for such collisions. $V_{n}^{\prime} $ is the neutral turbulent velocity at $L^{\prime} $. {\em Li and Houde} (2008) showed how $V_{n}^{\prime} $ and $L^{\prime}$ can be
derived from the position-position-velocity data cube with, most importantly, the line-width
difference between coexistent ions and neutrals. In this way, $L^{\prime}$ is estimated to be on the order of a {\em few hundred AU} (about 1 milli-pc), the size of a typical protostellar disc. So TAD must, to some degree, play a role in mitigating the magnetic braking catastrophe if disc angular momentum originates from turbulence.
The most direct way to test the above TAD model (Figure 12) is to probe turbulent velocity spectra of HCO$^{+}$ and HCN below the TAD scale, where the model predicts that
the ions should have a steeper slope. With the SMA, we have observed turbulent
velocity spectra down to the scale of 0.01 pc, where ions and neutrals still share the same slope.
With ALMA, it will be possible to probe turbulent velocity spectra well below 1 milli-pc. This also
allows direct measurements of $V_{n}^{\prime} $ and $L^{\prime}$, which can be used to estimate B-field strength with the above equation. 

While both models try to explain the ion/neutral line-width difference, the model discussed here ({\em Li and Houde} 2008) is independent from the model introduced earlier by {\em Houde et al.} (2000). {\em Li and Houde} (2008) relate line-width difference to $B_{pos}$ through turbulence spectra and the decoupling scale ($L^{\prime}$). These turbulence properties are not treated in the earlier model, which relates the line-width difference to the B-field orientation. The two models are sometimes confused in literature (e.g., {Crutcher} 2012). The observation that the coexistent ions and neutrals should have different turbulent velocities (line
widths) is also supported by simulations from various groups (e.g., {\em Tilley and Balsara} 2010, 2011; {\em Li, McKee and Klein} 2012). Though these
simulations can reproduce the ion/neutral line-width difference, the detailed results,
e.g., decoupling scale and slopes of ion/neutral turbulent velocity spectra, vary from code to code.
TAD simulations are challenging; all of these codes must either adopt some kind of approximation
or sacrifice resolution in order to realize a simulation within a reasonable CPU time.

 \bigskip
\noindent
\textbf{Observation XI: Dispersion of Magnetic Fields and the Turbulent Power Spectrum}
\bigskip

The aforementioned turbulent ambipolar diffusion decoupling scale
between the magnetic field (ions) and the neutral component of
the gas can also be inferred through polarization maps. This is accomplished
by the application and generalization of techniques of analysis well
known in turbulence studies to polarization data. More precisely,
something akin to a structure function of the second order often applied
to turbulent velocity fields ({\em Frisch 1995}) can be transposed to polarization
angles ({\em Dotson et al. 1996, Falceta-Gon\c{c}alves et al. 2008, Hildebrand et al. 2009, Houde et al. 2009, Houde et al. 2011 and Houde et al. 2013}).
That is, if we define by $\Delta\Phi\left(\ell\right)$ the difference
between two polarization angles separated by a distance $\ell$ on
a polarization map, then we can introduce the angular dispersion function
$1-\left\langle \cos\left[\Delta\Phi\left(\ell\right)\right]\right\rangle $,
where the average $\left\langle \cdots\right\rangle $ is performed
on all pairs of polarization angles for a given distance $\ell$.
It is straightforward to verify that the dispersion function is closely
related to the structure function of the polarization angles $\left\langle \Delta\Phi^{2}\left(\ell\right)\right\rangle $
through 

\[
1-\left\langle \cos\left[\Delta\Phi\left(\ell\right)\right]\right\rangle \simeq\frac{1}{2}\left\langle \Delta\Phi^{2}\left(\ell\right)\right\rangle 
\]

\noindent when $\Delta\Phi\left(\ell\right)\ll1$  ({\em Houde et al. 2013}). Although for the
purpose of the following discussion the dispersion or structure functions
could be used interchangeably, the advantage in using the dispersion
function is its close connection to the power spectrum (see below). 

With the assumption that the magnetic field can be modeled as the
sum of turbulent ($\mathbf{B}_{\mathrm{t}}$; i.e., zero-mean and
random) and ordered ($\mathbf{B}_{0}$) components, it is easy to
show that the structure (or the dispersion) function is also the sum
of corresponding turbulent and ordered functions  ({\em Houde et al. 2013}). That is,

\[
\left\langle \Delta\Phi^{2}\left(\ell\right)\right\rangle =\left\langle \Delta\Phi_{\mathrm{t}}^{2}\left(\ell\right)\right\rangle +\left\langle \Delta\Phi_{0}^{2}\left(\ell\right)\right\rangle .
\]

Given that the turbulent and ordered components of the field are usually
characterized by different length scales, with the ordered component
being of a larger scale, they can be cleanly separated \emph{without
any assumption on the exact morphology of the ordered (large-scale)
magnetic field}. This is a significant improvement on previous analyses
of polarization maps, where the determination of the magnetic field
dispersion (to be used, for example, in the so-called Chandrasekhar-Fermi
equation; {\em Chandrasekhar and Fermi} 1953) necessitated modeling the large-scale
magnetic field with a particular shape (e.g., an hourglass). Since
it is highly likely that this description of the large-scale magnetic
field with a predefined function will not, in general, perfectly fit
the true nature of the field, any error thus introduced in the analysis
will be propagated in subsequent calculations. 

An example of a dispersion function analysis applied to low-mass star
formation simulations taken from {\em Hennebelle et al.} (2011) is shown
in Figure 13. It is shown how the turbulent component of the
dispersion function (i.e.,
the turbulent autocorrelation function; symbols, bottom panel) can be extracted from the data (symbols, top
panel) by simply fitting the component due to the 
large-scale magnetic field by a Taylor series in a range where we
expect contributions from the turbulent magnetic field to vanish.

 \begin{figure}[ht!]
 \epsscale{1.2}
 \includegraphics[trim=1cm 0cm 1cm 0cm, clip=true, totalheight=0.24\textheight, angle=0]{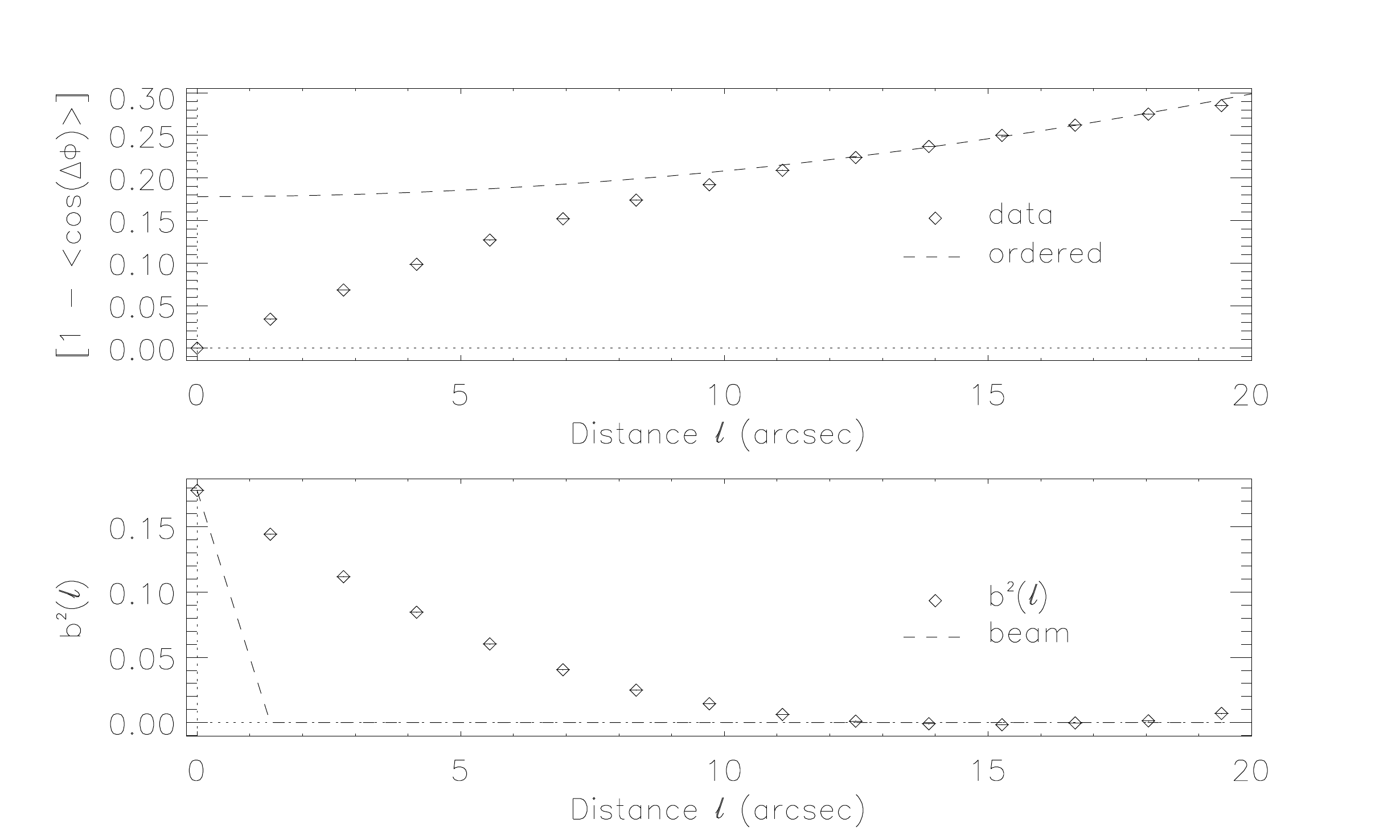}
 \caption{\small  Angular dispersion function of a low-mass star
formation simulation taken from {\em Hennebelle et al.} (2011). Top: the
total dispersion function (i.e., the sum of the turbulent and ordered
components) is represented by the symbols, while a Taylor series fit
to data located at $\ell>13\arcsec$ (broken curve) accounts for the
contribution of the ordered component to the total dispersion function
(shifted up by the constant level of the turbulent component in the
fitting range). Bottom: subtraction of the data from the Taylor series
fit (from the top panel) yields the normalized magnetic field autocorrelation
function (symbols), which is broadened by the finite resolution of
the data (i.e., the beam size; broken curve).
 }  
 \end{figure}

It is important
to note that the overall width of the autocorrelation function results
not only from the intrinsic correlation length of the magnetized turbulence
but also from the size of the beam defining the resolution of the
data. Still, the excess of the autocorrelation width beyond the beam\textquoteright{}s
contribution is a measure of the turbulence correlation length. Although
the problem is further complicated by the fact that the signal is
integrated through the line of sight and across the area subtended
by the telescope beam, it was shown by {\em Houde et al.} (2009) that the turbulent
correlation length and the relative level of magnetic turbulent energy can be readily determined when the underlying
observations are realized at high enough spatial resolution (i.e.,
when the beam width is narrower than the width of the turbulent
autocorrelation function, as is the case in the bottom panel of Figure 13). The knowledge
of these parameters is especially important to the application of
the Chandrasekhar-Fermi equation for estimating the magnetic field
strength $B_{\mathrm{pos}}$. This is because it then becomes possible
to account for the number of turbulent cells contained within the
column of gas probed by the observations and correct for the aforementioned
signal integration that averages down the turbulent dispersion function.
For example, {\em Houde et al.} (2009) performed a dispersion analysis on
a 350-$\mu$m SHARP map of the OMC-1 molecular cloud ({\em Vaillancourt et al.} 2008).
They determined the correlation length of the magnetized turbulence
to be 16 mpc and the number of turbulent cells probed by the telescope
beam to be approximately 21, yielding a plausible value of $B_{\mathrm{pos}}=760\;\mu$G.
CN$\left(N=1\rightarrow0\right)$ Zeeman measurements yielded a
line-of-sight field strength of approximately 360 $\mu$G in this
source; see {\em Crutcher et al.} (1999).

Being the inverse of the spectral width of the turbulent power spectrum,
the turbulent correlation length is to some extent linked to the turbulent
ambipolar diffusion decoupling scale. As was mentioned in the previous
section, this is the scale of turbulent energy dissipation that characterizes
the upper end (in $k$-space) of the power spectrum. This spectrum
is connected to the dispersion function through a simple Fourier transform

\[
\left\langle \mathbf{\overline{B}}_{\mathrm{t}}\cdot\mathbf{\overline{B}}_{\mathrm{t}}\left(\ell\right)\right\rangle \Leftrightarrow R_{\mathrm{t}}\left(k\right)\left\Vert H\left(k\right)\right\Vert ^{2}
\]

\noindent where $\left\langle
  \mathbf{\overline{B}}_{\mathrm{t}}\cdot\mathbf{\overline{B}}_{\mathrm{t}}\left(\ell\right)\right\rangle$
is the measured autocorrelation function of the turbulent magnetic field,
$R_{\mathrm{t}}\left(k\right)$ is the intrinsic turbulent
power spectrum, and $H\left(k\right)$ is the spectral profile of the
telescope beam. The turbulent power spectrum associated with the autocorrelation
of Figure 13 is shown in Figure 14.
Multiplying the resulting power spectrum (symbols, top panel) by $2\pi k$
yields a one-dimensional spectrum (solid curve), which, for example,
can be fitted to a power law in its inertial range (broken curve).
In this example, the log-log plot of the one-dimensional spectrum
(symbols, bottom panel) reveals a (numerical) dissipation scale at
$\log\left(k/2\pi\right)\simeq-0.7$ arcsec$^{-1}$. {\em Houde et al.} (2011) have applied
this spectral analysis to three high resolution polarization maps
obtained with the SMA (for Orion KL, IRAS 16293, and NGC 1333 IRAS
4A) and although the spectra obtained are not as cleanly resolved
as the one shown in Figure 13, a dissipation scale
was clearly detected at approximately 10 mpc for Orion KL. This scale
is most likely associated to turbulent ambipolar diffusion, as discussed
in the previous section and first detected by {\em Li and Houde} (2008) in M17
using comparisons of HCO$^{+}$ and HCN line widths. 

  \begin{figure}[ht!]
 \epsscale{1.05}
\includegraphics[trim=.9cm 0cm 1cm 0cm, clip=true, totalheight=0.23\textheight, angle=0]{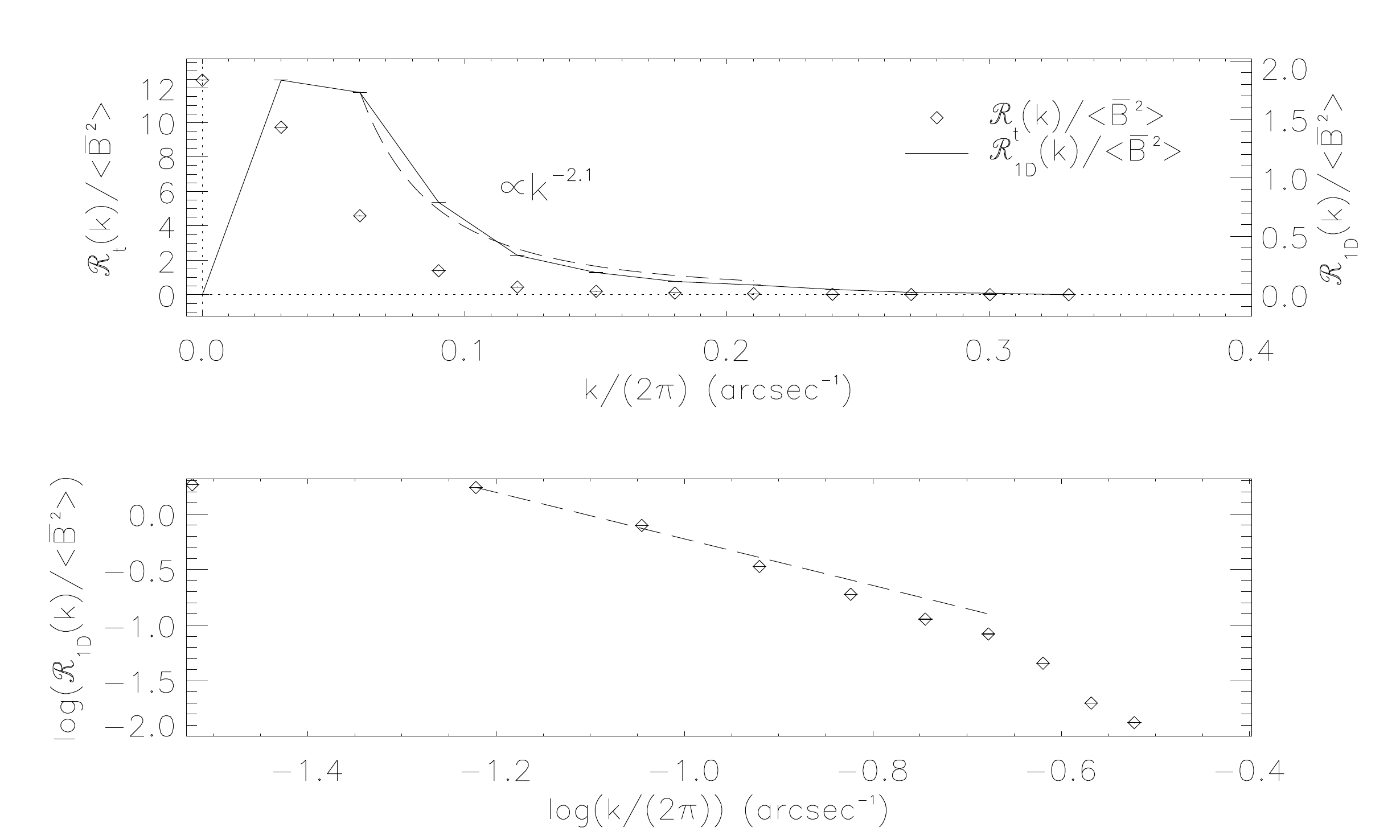}
 \caption{\small  The turbulent power spectrum (symbols) calculated
from the Fourier transform of the autocorrelation function shown in
the bottom panel of Figure 13; this spectrum is
multiplied by $2\pi k$ to obtain the one-dimensional (Kolmogorov-type)
power spectrum (solid curve). Bottom: Log-log plot of the one-dimensional
turbulent power spectrum (symbols). Both panels show a power law fit
to the inertial range of the one-dimensional turbulent power spectrum
(broken curve); the (numerical) dissipation scale is seen in the bottom
panel at $\log\left(k/2\pi\right)\simeq-0.7$.
 }  
 \end{figure}

\bigskip
\noindent
\textbf{Observation XII: Alignment of Outflows and Flattened Infall Envelopes with Cloud
B-fields?}
\bigskip

Observationally, we have not yet fully characterized the relationship between the
symmetry axes of structures in the protostellar environment, such as bipolar
outflows and flattened infall envelopes, and the orientation of the local cloud
B-field.  However, new instrumentation is improving the situation, leading to new
tests for disk formation models such as those reviewed at the beginning of section
5.   If B-fields in cloud cores are sufficiently strong then one expects collapse to
proceed primarily along field lines, leading to infall envelopes that are flattened
along the direction of the B-field (Observation VIII).  Such magnetically-shaped flattened infalling
gas structures (a.k.a. ``pseudodisks"; {\em Galli and Shu} 1993; {\em Allen, Li and Shu} 2003) have
characteristic scales of several thousands of AU for low-mass protostars.  If cloud
cores have their rotation axes aligned with the B-field due to magnetic braking
({\em Mouschovias and Paleologou} 1979) then the smaller scale Keplerian disks should also
have their symmetry axes parallel to the B-field.  We can test for such an alignment
by comparing B-field orientation with outflow axis.  Not all simulations of
magnetized core collapse predict good alignment between core B-field and Keplerian
disk axis.  For example, see {\em Joos et al.} (2012).  

Tests for alignment between protostellar structures and B-fields have employed
various techniques; here we restrict the discussion to submm or mm-wave polarimetry.
 {\em Curran and Chrysostomou} (2007) carried out submm polarimetry using SCUBA/JCMT for a
sample of 16 high-mass star forming regions and compared mean B-field directions
with outflow axes.  The projected misalignment angles were nearly uniformly
distributed between 0 and 90 degrees, suggesting the absence of any correlation, as
confirmed with a Kolgomorov-Smirnov test yielding p = 0.849.  The authors pointed
out that for outflows lying nearly parallel to the line of sight it would be
difficult to observe any intrinsic alignment that might exist between outflow and
field. However, when they omitted from
their analysis all outflows having inclination angles below 45 degrees (i.e., those
oriented relatively close to the line of sight), the distribution was still
consistent with a random one (p = 0.572).

\bigskip 
 \begin{figure*}
  \includegraphics[trim=.4cm .2cm 0cm 0cm, clip=true, totalheight=.55 \textheight, angle=0]{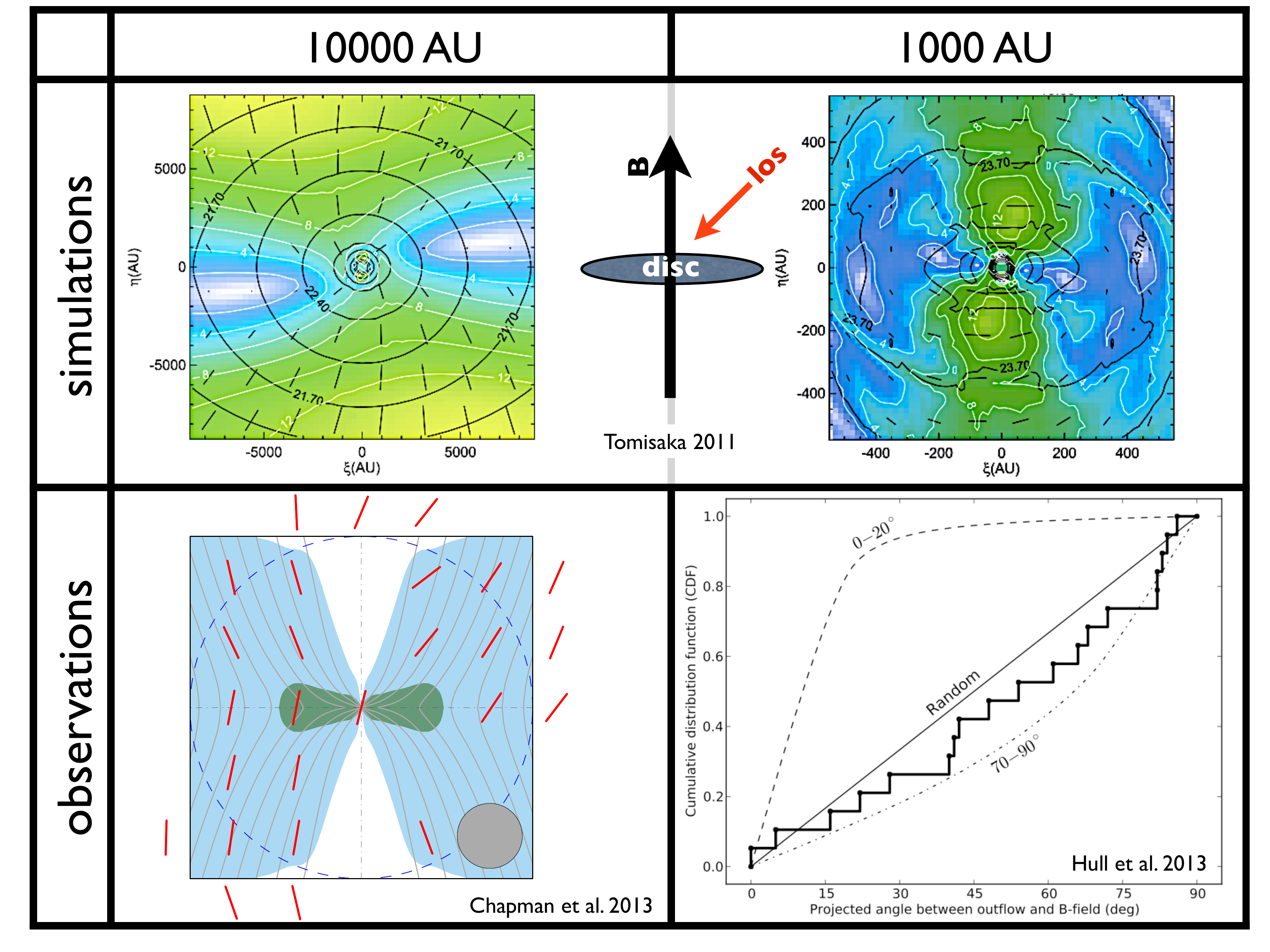}
 \caption{\small {\em Simulation-} B field directions inferred from the polarization of thermal dust emission ({\em Tomisaka} 2011) based on a MHD and radiation-transfer simulation. The simulation results in a disc perpendicular to the initial B field, and bipolar outflows driven by magnetic pressure. The figure shows the projections along a line of sight {\em $45^{\circ}$} from the initial B field direction. The relative column density is shown by the white contours and the colors. The dark vectors and contours show the field directions and polarization degrees. The {\em left panel}, on a scale comparable to {\em Chapman et al.} (2013), shows field directions roughly perpendicular to the pseudodisk. The {\em right panel} is a zoom-in of the left panel and shows the central 1000-AU region, which is comparable to {\em Hull et al.} (2013). The outflows are visible (the greenish zone in the center). Most importantly, the field morphology on 1000-AU scale starts to be affected by the disc rotation and can be oriented at any direction; i.e., {\em the mean field direction on this scale has nothing to do with the mean field direction of the cloud core}. 
  \endgraf
  \
 \endgraf
 {\em Observations-} 10000 AU: Source-average observed B-field directions (red bars) superposed on model
magnetic field lines (grey) and gas density contours (green and blue shading),
adapted from {\em Chapman et al.} (2013). The model is from {\em  Allen, Li and Shu} (2003).  The observed
source-average B-field map has a mean direction that is nearly vertical and shows
hints of a pinch, in accordance with the model.  1000 AU: On the other hand, a B field survey from {\em Hull et al} (2013) shows no correlation between the mean field and the inferred disc orientation.
} 
\end{figure*}

{\em Wolf et al.} (2003) and {\em Davidson et al.} (2011) carried out similar types of
observational studies for the case of nearby, low-mass cores, pointing out that such
targets have generally simpler geometries, are often more isolated, and most
importantly are the types of objects most often simulated by theorists.  {\em Wolf et al.}
(2003) used SCUBA/JCMT polarimetry to map four targets and {\em Davidson et al.} (2011)
used the SHARP submm polarimeter at CSO ({\em Li et al.} 2008b) to study three sources.  Both results are suggestive of an alignment between field and outflow, at least for the
younger sources, but neither study reported statistical tests for an overall
alignment.  Two very recent studies of nearby low-mass cores that did carry out such
tests are SHARP/CSO polarimetry by {\em Chapman et al.} (2013) on 10,000 AU scales, and
CARMA interferometric 1.3 mm polarimetry by {\em Hull et al.} (2013) on 1000 AU scales. 
{\em Chapman et al.} (2013) restricted themselves to single Class 0 sources while {\em Hull et
al.} (2013) included multiples as well as Class I targets.  

The survey of {\em Chapman et al.} (2013) included seven targets, and they compared
position angles of mean B-field, flattened infall envelope (``pseudodisk"), and
outflow.  Following the example of {\em Curran and Chrysostomou} (2007), they made use of
the estimated outflow inclination angles.  They found a tight alignment between the
position angles of pseudodisk minor axes and outflow axes, and they exploited this
by using the outflow inclination angles as proxies for the unknown pseudodisk
symmetry axis inclination angles.  Using all these angles, they tested for
correlations, in 3-dimensional space, between B-field and pseudodisk symmetry axis
and between B-field and outflow axis, finding evidence for positive correlations in
both cases, with 95\% and 96\% confidence respectively.  As a further test, they
combined polarimetry data from the subset of their sources having inclination angle
below 45 degrees, first rotating each map according to its apparent pseudodisk axis,
to form a source-averaged magnetic field map having improved signal-to-noise ratio. 
This map, shown in Figure 15, shows that the overall direction of the source-averaged
B-field is roughly perpendicular to the pseudodisk plane, as expected from their
reported correlation between pseudo-disk symmetry axis and B-field.  Similar results
were obtained when the individual maps were rotated according to their outflow
position angles rather than referencing to the pseudodisks.

The interferometric polarimetry survey of {\em Hull et al.} (2013) included 16 sources,
and they compared the position angles of mean B-field and outflow.  The projected
misalignment angles were found to range from 0 to 90 degrees with a nearly uniform
distribution, consistent with a random alignment.  The Kolgomorov-Smirnov test
yielded p = 0.64.  Inclination angles were not considered.  

How can we reconcile the {\em Chapman et al.} (2013) and {\em Hull et al.} (2013) surveys?  One
possibility is that the inclusion in the CARMA survey of the (more complex)
multiples and (more evolved) Class I sources obscures the correlation.  This is
suggested by the fact that the {\em Hull et al.} (2013) results do not seem random if we
consider only the four relatively simple and young sources that are included in both
surveys.  L1157 and L1448-IRS2 show very good alignment between field and
outflow/pseudodisk in both surveys, and Serp-FIR1 can be discounted as its outflow
lies nearly parallel to the line of sight.  This leaves L1527, where on the scales
of the SHARP/CSO maps the B-field is consistent with a pinched poloidal field
aligned with the outflow axis, while on the smaller scales probed by CARMA the field
is perpendicular to the outflow, suggesting a transition of field directions from large to small scales.
This transition is demonstrated in numerical simulations of discs and outflows (e.g., Figure 15).
The simulation in Figure 15 ({\em Tomisaka} 2011) has the core B-field, disc minor axis, and outflow well aligned, and the B-field is perpendicular to the disc on 10,000 AU scales (left panel) but can be at any direction on 1000 AU scales (right panel). This is consistent with both surveys. Zooming in to the central 300 AU scale, {\em Tomisaka} (2011) sees toroidal fields. Because of this field complexity at small scales due to outflows and disc rotation, B-field directions on 1000 AU scales are not representative of core fields and thus cannot put constraints on whether cloud fields and discs should be aligned (e.g., {\em Tomisaka} 2011) or unaligned (e.g., {\em Joos et al.} 2012).


The morphology of gas density and B-field evident in Figure 15 is suggestive of
B-fields guiding protostellar collapse, just as was the hourglass B-field discovered
earlier in the Class 0 protobinary NGC 1333 IRAS 4A by {\em Girart, Rao and Marrone}
(2006).  However, we should not discount the possibility that feedback may influence
the appearance of such maps.  Is the field controlling the collapse or visa versa? 
In the case of NGC 1333 IRAS 4A, the SHARP/CSO map of {\em Attard et al.} (2009) shows
that B-fields continue smoothly to larger scales (see the polarization vectors in Figure 17), indicating that indeed they guide the collapse in this source.  But for the {\em Chapman et al.} (2013) result of Figure 15
data on larger scales remains to be compiled.  In this case near-IR polarimetry
({\em Clemens et al.} 2007) may provide a way forward.  In the longer run, large area
submm polarimetry from the stratosphere ({\em Dowell et al.} 2010; {\em Pascale et al.} 2012)
will be combined with ALMA polarimetry to yield multi-scale maps for scores of
targets.  

Improvements will also come from the development of new analysis techniques.  One
troublesome effect is the ``bias toward 90 degrees''.  To understand this, consider
taking the average of a set of measured B-field position angles lying in the
interval 0 to 180 degrees.  If the distribution is broad, there will be a tendency
for this average to lie closer to 90 degrees rather than to 0 degrees or 180
degrees.  For example, note that all 16 of the well-separated independent clouds
studied by {\em Curran and Chrysostomou} (2007) have mean B-field angles lying between 45
and 135 degrees.  The strength of this bias varies from study to study, but it is
always present unless removed.  A simple method for removing it is to use the Equal
Weight Stokes Mean statistic ({\em Li et al.} 2006) in place of a straight mean.


 \bigskip
\centerline{\textbf{ 6. DISCUSSION}}
\bigskip

B-field observations are difficult and undoubtedly involve many uncertainties. However, one should not give up hope, because many useful arguments can still be made despite the uncertainties. For example, Observation I properly treats the ambiguity of the CO line polarization direction. Also, although there is no way to precisely control the region probed by optical polarization data, contamination from the lines of sight should not ``create" the correlations seen in Figures 6 \& 8; if anything, the contamination does the opposite. Many factors indeed contribute to emission line widths, but how many are even arguably the possible cause of the ion/neutral line-width difference that is correlated with B-field strength  (Observation X)?   

In the following, we discuss critically some of the ''evidence'' that has been used to argue against the strong-field scenario: 

1. Data scattering under the magnetic critical line in the B$_{los}$-vs.-$N_{H}$ plot (Figure 4) implies super-Alfv\'enic turbulence?

2. The $\sim2/3$ slope in the B$_{los}$-vs.-$n_{H}$ plot implies isotropic core contraction?

3. The optical polarization angle dispersions (Figure 6 \& 8) are too large for dynamically dominant B fields?

4. Sub-mm ``polarization holes'' imply that high-density cores are not probed by sub-mm polarimetry? 

5. Observational evidence for super-Alfv\'enic turbulence?

\bigskip 
{\bf 6.1 Zeeman measurement scattering and mass-to-flux ratio}

The four cores and envelopes in Observation IX (from Perseus and Taurus molecular clouds) are mostly within regions with A$_{V} > 4$ mag, which is very likely to be higher than the critical column density (see Figure 4). The cloud contraction thresholds defined by the observed column density PDFs of these two clouds ({\em Kainulainen et al.} 2009) are indeed lower than A$_{V} = 4$ mag. Thus, it is possible self-gravity forms the four cores without help from either ambipolar diffusion or super-Alfv\'enic turbulence. This agrees with the suggestion from {\em Elmegreen} (2007) that core evolution starts close and stays close to the magnetically critical state. Whether these cores can further fragment with or without ambipolar diffusion is another story, because sub-regions have smaller mass-to-flux ratios (mass $\propto r^{3}$ and flux $\propto r^{2}$, where $r$ is the scale). 

A criticism of this magnetic-criticality-hypothesis is that data points in Figure 4 are consistent with B$_{total}$ widely scattered under the critical line, not just slightly below it  ({\em Crutcher et al.} 2010), and this is seen in simulations assuming weak B fields.  
However, with the lessons learned from many of the observations discussed here, we should review the way the data scattering in Figure 4 is interpreted.


First, the efforts to explain Observation IX improved our understanding of Zeeman measurements. While the interpretation of  Observation IX had been very controversial (e.g., {\em Mouschovias and Tassis} 2009; {\em Crutcher et al.} 2010), various groups (e.g.,  {\em Lunttila et al.} 2008, {\em Bertram et al.} 2012, {\em Mouschovias and Tassis} 2009, {\em Bourke et al.} 2001) seem to reach one agreement: B$_{los}$ reversals will cause Zeeman measurements to {\em underestimate B$_{los}$}, which can explain Observation IX.

Second, mass contraction along B fields (Observation II \& VIII) will not increase the total mass in the system, but will increase the {\em column density} observed, as long as the line of sight is not aligned with the B fields ({\em Li et al.} 2011). 

Both of these effects will increase column density-to-B$_{los}$ ratios and should be considered when interpreting data in Figure 4. Given the same intrinsic B-field strength, the same total mass and even the same non-zero offset between the line of sight and the mean B-field, the appearance of Figure 4 can still vary. Turbulence brings a data point {\em downward} by dispersing (not necessarily tangling; see Observation IX) the B-field direction and gravitational contraction/turbulence compression channeled by B-fields moves a data point toward the {\em right} in Figure 4. Observing an oblate core with B-field direction close to the symmetry axis, as suggested by Observation VIII, with different lines of sight also situates data points at different positions in Figure 4: while B$_{los}$ certainly moves down as the line of sight moves away from the core axis, also note that the column density will increase as our view becomes more edge-on to the oblate core. All these effects will cause column density-to-B$_{los}$ ratios to {\em overestimate} mass-to-flux ratios and cause the data points in Figure 4 to be scattered {\em under} the magnetically critical line. So even if B-fields are dynamically dominant, data points should still be scattered under the critical line.
For the same reason, ''R'' is scattered in Figure 11.

\bigskip 
{\bf 6.2 Observations VIII versus the 2/3 slope of the B$_{los}$-vs.-$n_{H}$ plot}

Observation VIII also supports the idea that core evolution should stay close to the magnetically critical state ({\em Elmegreen} 2007). If the cores are highly super-critical, gravitational contraction should be isotropic and their shape should not correlate with B field directions as seen in Observation VIII. However, {\em Crutcher et al.} (2010) concluded that a dynamically important B field during core formation is inconsistent with the Zeeman measurements from clouds and cloud cores (based on OH and CN data respectively), because the upper limit of the B$_{los}$-vs.-$n_{H}$ plot has a slope $\sim2/3$. This slope implies that gravitational contraction is isotropic. 
One possible explanation for this discrepancy is that the OH measurements are from a dark cloud survey while the CN data is mostly from massive cluster forming clumps in GMCs. Since it is unlikely for dense cores of nearby dark clouds to evolve into massive cluster-forming clumps, using the slope to infer an isotropic collapse is questionable. 

Also worth noticing is that isotropic contraction should result in radial-like B-field morphologies. However, this is rarely observed (e.g., see Figures 5D, 6, 10 \& 17). Even the ``hour-glass" shaped field morphologies (e.g., Figure 17) are not often observed .  

%

\bigskip
{\bf 6.3 Optical polarization direction angle dispersion}

The orderliness of B fields can be used as an indication of the relative strength between B fields and turbulence. Based on numerical MHD simulations, field dispersion is sensitive to the Alfv\'enic Mach number ($M_{A}$; Figure 16); even slightly super-Alfvenic turbulence is enough to significantly distort or even randomise (Figure 16) field directions. The telescope beam average may lower the dispersion (Observation XI), but cannot artificially introduce a well-defined mean field direction to a random field. So the fields shown in Figures 6, 7 \& 8 cannot be super-Alfv\'enic (random). Moreover, the projected 2D B-field dispersion is usually larger than the intrinsic 3D field dispersion (Figure 16).   

   \begin{figure}[ht]
 \epsscale{1.05}
\includegraphics[trim=0cm 0cm 0cm 0cm, clip=true, totalheight=0.35 \textheight, angle=0]{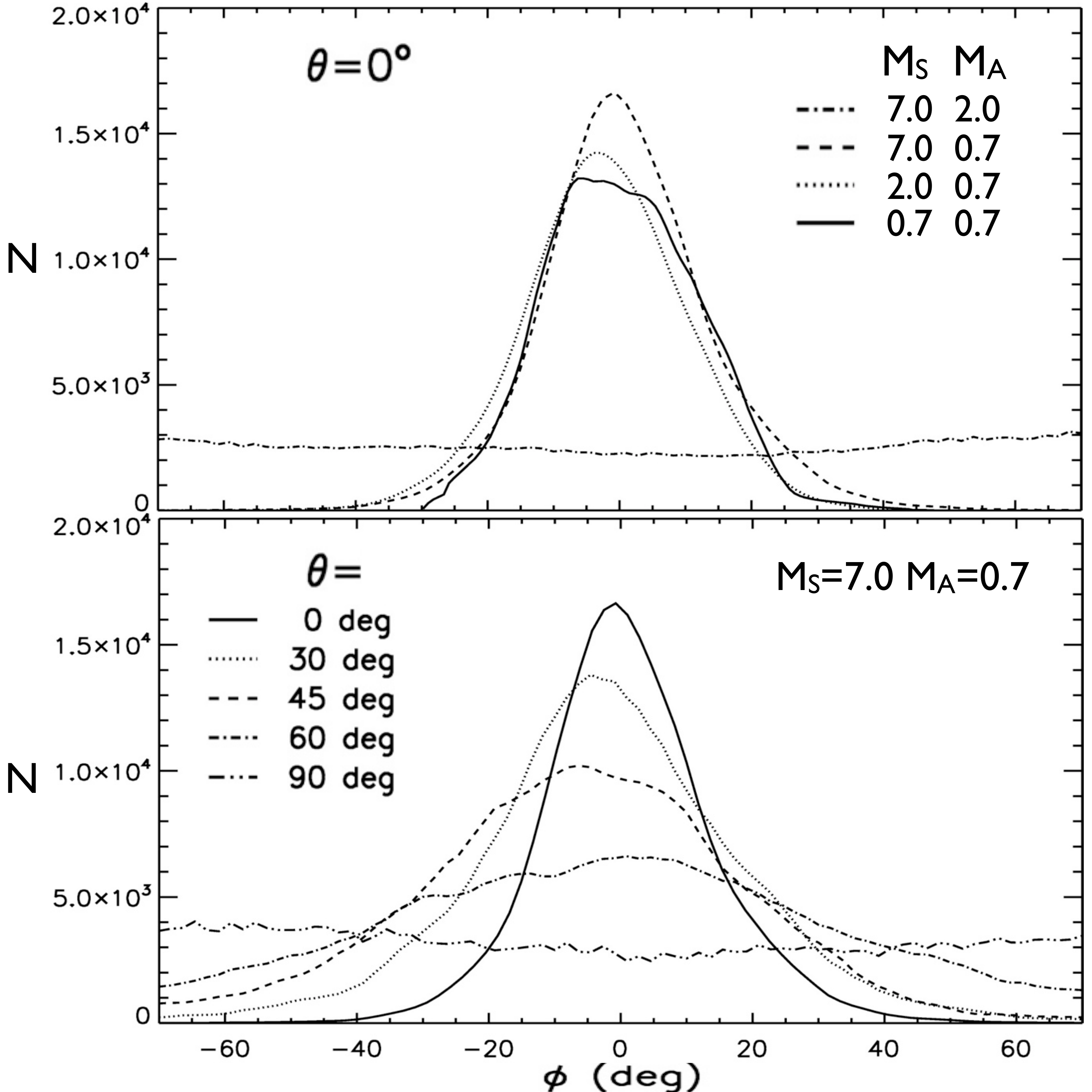}
 \caption{\small  Simulated sub-mm polarization direction angle ($\phi$) dispersion ({\em Falceta--Gon\c{c}alves et al.} 2008). $\theta$ is how far the line of sight is from being perpendicular to the mean field direction; $M_{S}$ and $M_{A}$ are sonic Mach number and Alfv\'enic Mach number. The {\em upper panel} shows that super-Alfv\'enic ($M_{A} > 1$) turbulence results in random polarization direction. The {\em lower panel} illustrates the projection effect: $\phi$ dispersion increases with $\theta$.
 }  
 \end{figure}
 
Following {\em Chandrasekhar and Fermi} (1953), we can estimate the lower limit of B-field direction dispersions for super-Alfv\'enic turbulence and compare this with the dispersion (STD) shown in Figures 5 \& 7 (on average 31 degrees). Assuming that the STD of B-field direction ($\sigma$, observed with a line of sight perpendicular to the mean field) is completely due to gas turbulence, Chandrasekhar and Fermi derived the relation (CF relation) : $\sigma (radians) = [4\pi\rho]^{1/2}$v/B, where B, $\rho$ and v are, respectively, the B-field strength ($Gauss$), density ($gm/cm^{3}$) and line-of-sight turbulent velocity ($cm/s$). They used the small angle approximation, and numerical simulations (e.g., {\em Ostriker et al.} 2001) indicate that the CF relation is a good approximation only when $\sigma < 25^{\circ}$.  The average optical polarization STD is $31^{\circ}$, which is too large for using the CF relation. {\em Falceta--Gon\c{c}alves et al.} (2008) improved the relation by replacing $\sigma$ with tan($\sigma$), and numerically showed that this new relation is applicable for larger $\sigma$.  Setting $M_{A}$=1, i.e., v = B/$[12\pi\rho]^{1/2}$, the improved CF relation gives $\sigma = 30^{\circ}$, which is almost identical to the observed value. 

However, it is not only $M_{A}$ that affects the observed angle dispersion. Projection effects (Figure 16) and non-turbulent structures of the B field also contribute to the angle dispersion. Non-turbulent B field structures include, e.g., those caused by stellar feedback and Galactic B field structures (inter-cloud B fields in Figures 6 \& 8 spread out over hundreds of pc along a line of sight). Observation XI introduced new analytical tools for removing non-turbulent structures in the polarimetry data, which has however not been applied to the optical polarization data. With all these non-turbulent factors that can significantly increase the B-field dispersion, the B-field dispersion ($31^{\circ}$) is only nearly equivalent to the trans-Alfv\'enic condition assuming turbulence is the only force that can deviate field directions. This suggests that the turbulence is sub-Alfv\'enic.

\bigskip 
{\bf 6.4 Sub-mm ``polarization holes''}

It is generally observed that the degree of polarization from thermal dust emission decreases with increasing column density. An example is shown in Figure 17. This phenomenon is called ``polarization holes''. A straightforward and popular interpretation is that increasing density will reduce the dust grain alignment efficiency; {\em Padoan et al.} (2001), for example, suggested that grains are not aligned for $A_{v} > 3$ mag. This simple scenario however fails to explain the other general observation that was found later with interferometers: zooming in onto polarization holes with interferometers reveals polarization {\em in a similar direction but with much larger polarization fractions}, some even larger than those from the low-density sight lines in the single-dish map (Figure 17). Since interferometers filter out regions that are less dense and more extended, lower polarization should have been observed by interferometers if grain alignment is ``turned off" in high-density regions.

Assuming constant grain alignment efficiency, increasing dispersion of B-field direction angles can also lower the polarization fraction. Whether this idea can explain polarization holes depends on whether B-field direction dispersion increases with column density. This seems plausible, because, for a typical clump as shown in Figure 17, larger column density usually implies larger line-of-sight dimensions, which should host a larger velocity dispersion based on Larson's law. Then the CF relation implies that B-field direction dispersion is proportional to the velocity dispersion and, thus, column density (assuming field strength roughly proportional to the square root of the density). {\em Li et al.} (2009) indeed observed that polarization fractions decrease with increasing LOS velocity dispersions in DR21; the correlation is even higher than between polarization fraction/column density.

Here we test this idea using the simulations from {\em Ostriker et al.} (2001). First we select a subregion (Figure 18) of the beta=0.01 case which contains the densest filaments in the simulation. In the X-Y plane, for example, we study the correlation between column densities and B-field direction angle dispersions after averaging along Z. The result is shown in Figure 18 - a clear positive correlation is seen. Note that here no changes in grain alignment efficiency are involved. Simply the B-field dispersion along the line of sight is enough to cause polarization holes. 

 \begin{figure}[ht!]
 \epsscale{1.0}
\includegraphics[trim=0cm 0cm 0cm 0cm, clip=true, totalheight=0.33 \textheight, angle=0]{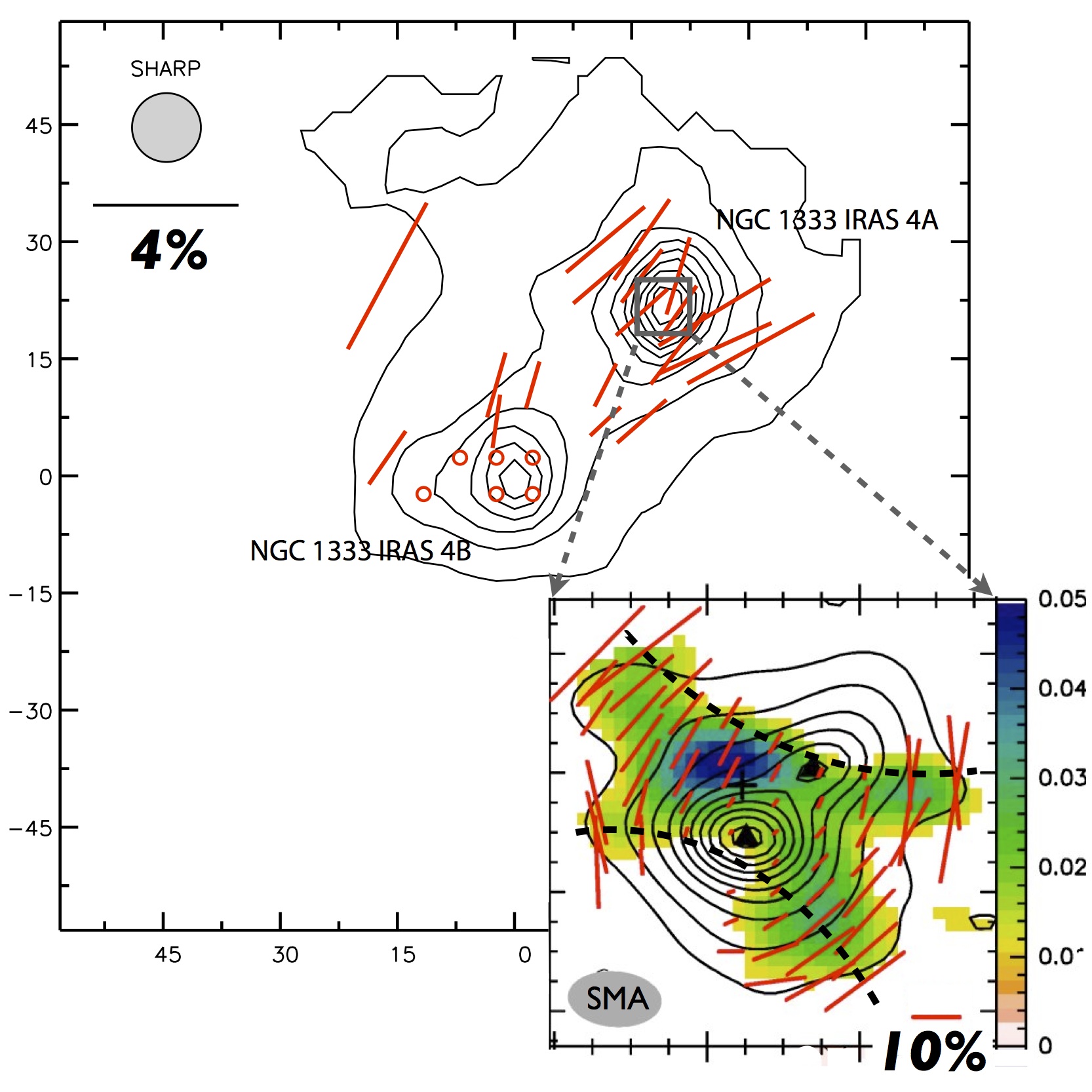}
 \caption{\small Low grain alignment efficiency cannot explain polarization holes.  {\em Upper panel}: The polarization holes in NGC 1333 observed by SHARP/CSO ({\em Attard et al.} 2009).  Dust emission flux is shown as contours. The vector lengths are proportional to the polarization fractions and the circles are where the polarization is too low to be detected by SHARP. In the upper-left corner are the single-dish beam size and a sample vector with the length of $4\%$ polarization. The coordinates are offsets in arcsec. {\em Lower panel}: {Girart et al.} (2006) used the SMA to focus on one of the polarization holes and revealed much larger degrees of polarization compared to the single-dish observation; note the sample vector of $10\%$ polarization in the lower-right corner. This certainly goes against the idea that higher density will lower the grain alignment efficiency. The synthesised SMA beam is shown in the lower-left corner. Dust emission flux is shown as contours and the color indicates polarized flux. 
Note that the vectors in both maps show the submm polarization directions. The inferred B-field directions are orthogonal to the vectors. The dashed lines in the SMA map trace two B-field lines, which appear as an hourglass shape. The B-field direction from the SHARP map is roughly aligned with the short axis of the core (see Observation VIII) and also aligned with the axis of the hourglass.
 }
 \end{figure}

 \begin{figure}[ht!]
 \epsscale{1.0}
\includegraphics[trim=0cm 0cm 0cm 0cm, clip=true, totalheight=0.55 \textheight, angle=0]{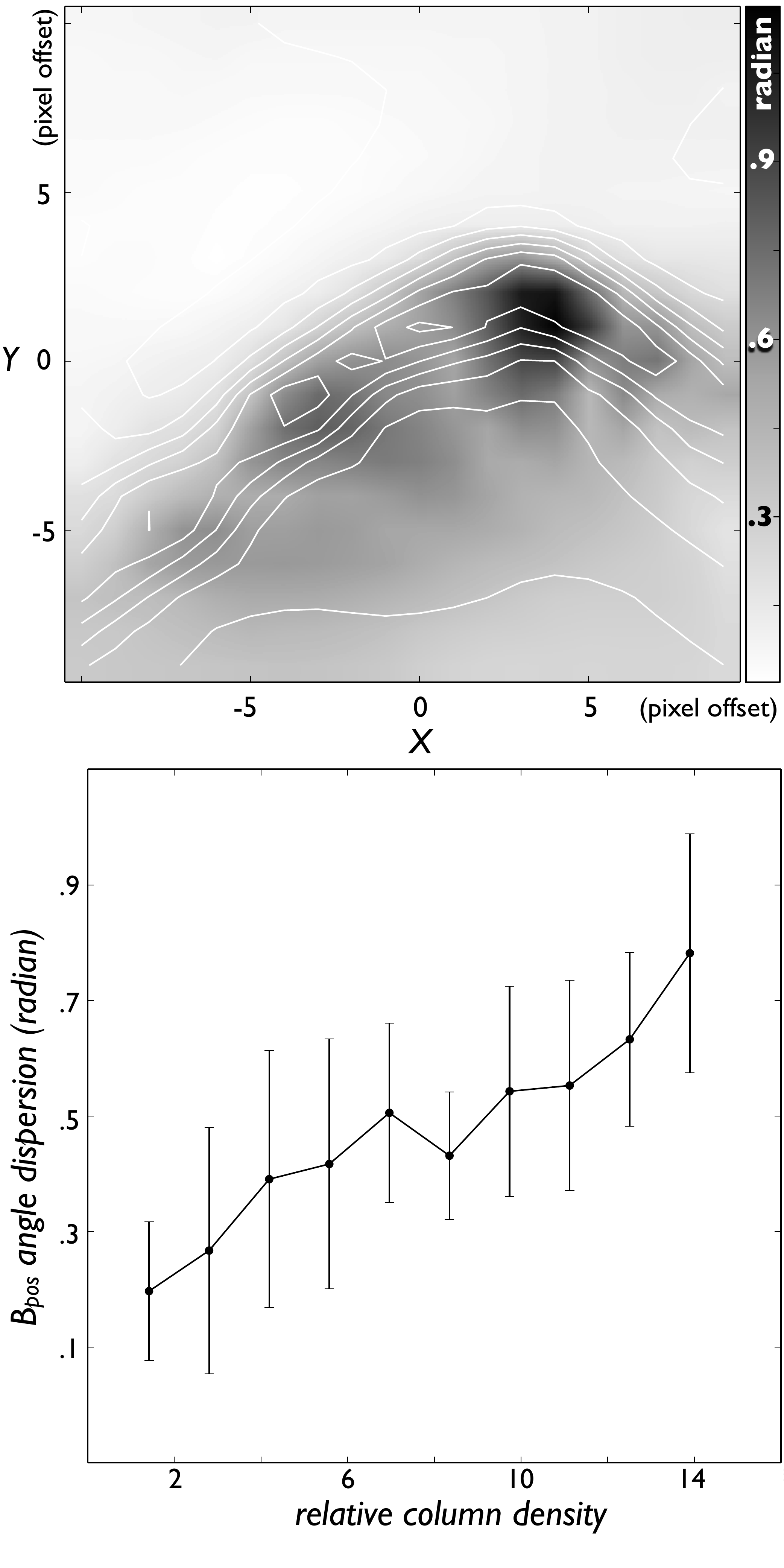}
 \caption{\small An illustration of how polarization holes may be caused by B-field structures in a sub-Alfv\'enic cloud. {\em Upper panel}: {\em Ostriker et al.} (2001) simulated a sub-Alfv\'enic ($\beta=0.01$) cloud and here we study the densest filament (pixel coordinates: x = 227-247; y = 50-70; z = 237-250) in this simulation . The contours show the mean density along each line of sight; the values of the  highest contour and the spacing between adjacent contours are respectively 140  and 20 times of the uniform density in the initial condition. The grayscale shows the $B_{pos}$ angle dispersion along each line of sight. $B_{pos}$ refers to the plane-of-sky (X-Y plane) component, which will affect the polarization. The higher the angle dispersion, the lower the polarization fraction, assuming constant grain alignment efficiency. {\em Lower panel}: $B_{pos}$ dispersion versus column density based on data in the upper panel. The $B_{pos}$ dispersion indeed increases with column density, and thus polarization holes at high column density are expected even with a constant grain alignment efficiency. 
 }  
 \end{figure}

\bigskip

{\bf 6.5 Observational support of super-Alfv\'enic turbulence?}

While the observations we reviewed point toward a picture of sub-Alfv\'enic turbulence, there are other observations which are used to support the scenario of super-Alfv\'enic clouds. Here we briefly discuss these observations and explain why these claims are based on assumptions which may not be fulfilled.
\bigskip

\noindent
\textbf{ Observation XIII: Power Spectra Indices of Cloud Column Densities}
\bigskip

One analysis commonly used in support for super-Alfv\'enic clouds is the column density power spectra suggested by {\em Padoan et al.} (2004). Their simulations show that Alfv\'enic flows provide a power law index for column density of 2.25, while highly super-Alfv\'enic clouds have an index around 2.7. Comparing with these simulations, they concluded that Perseus, Taurus, and Rosetta are all super-Alfv\'enic, because the power-law indices of their $^{13}CO$ maps are around 2.75. 

However, there are several caveats to the above conclusion. First, the simulated power-law index depends on the exact set-up of the simulations. For example, cloud simulations of {\em Collins et al.} (2012) show much shallower spectra of column densities, and the power-law indices are not correlated with magnetic Mach numbers. Second, the range of the observed power law indices is between 2 and 3; see Table I of {\em Schneider et al.} (2011), for example. Third, different tracers can result in very different power-law indices. For example, the indices of Perseus, Taurus, and Rosetta are, respectively, 2.16, 2.20, and 2.55 when dust extinction is used to trace the column densities ({\em Schneider et al.} 2011). {\em Schneider et al.} (2011) concluded that the indices probed by dust extinction are usually significantly lower than those probed by CO. With careful comparison of extinction, thermal emission and CO maps of Perseus, {\em Goodman et al.} (2009) concluded that dust is a better tracer of column density than CO, because it has no problems of threshold density, opacity, and chemical depletion.

Furthermore, as we have seen in Observation V, filamentary structures are equivalent to anisotropic autocorrelation functions (which is how the filament directions are defined) and, thus, anisotropic power spectra. The averaged index in this case depends on how the filamentary structure is projected on the sky. These considerations put in doubt the conclusion that empirical column density power laws support super-Alfv\'enic states of molecular clouds.

\bigskip
\noindent
\textbf{Observation XIV: Cloud/core B-field Direction vs. Galactic Disc Orientation}
\bigskip

As another argument against the strong-field scenario, {\em Stephens et al.} (2011) and {\em Bierman et al.} (2011) used the fact that cloud or core fields are not aligned with the Galactic disc to conclude that cloud B-fields must have decoupled from the Galactic B-fields. Their argument relies on the assumption that Galactic fields are largely aligned with the disc plane, which, however, is not the case at the scale of cloud accumulation length as we will show in the following.

Figure 6 in {\em Stephens et al.} (2011) shows an angle distribution of almost all the polarimetry detections from the {\em Heiles} (2000) catalog, and the distribution clearly peaks in the direction of the Galactic disc plane.  Note that this plot contains stars from distances of 140 pc to several kpc, and thus shows the (Stokes) mean B-fields from various scales because the polarization of a star samples the entire sight line (Figure 19). As a result, one cannot establish from their plot whether the B-field coherence happens at every scale or only at certain scale ranges. 

To distinguish between the two possibilities, in Figure 19, we plot similar polarization distributions but only for stars with distances within 100-pc bins centered at, respectively, 100, 300, 700, 1500, and 2500 pc in distance. We also use the optical data archive of {\em Heiles} (2000). We exclude data for which the ratio of the polarization level to its uncertainty is less than 2. The numbers of stars in each distance range are, from nearest to farthest, 1072, 339, 116, 82, and 51. At 100-pc scale the distribution is very flat, i.e., Galactic B-fields can have any direction. As shown in Figure 19, the so-called coherent Galactic B field only appears at scales above 700 pc, where structures at smaller scales are averaged out. 

Also shown in Figure 19 (with a dashed line) is the distribution of the B-field directions from 52 cloud cores at pc to sub-pc scales from {\em Stephens et al.} (2011). They concluded that the core B-fields must have decoupled from the Galactic B-fields, because the direction distribution of the core fields is not as peaked as their Figure 6. However, as the accumulation length of even a GMC is only $\sim$400 pc ({\em Williams, Blitz and McKee} 2000), the core B-fields are not expected to be related to Galactic B-field structures larger than 400-pc. In fact, the distributions of the core fields and the Galactic fields at 100-300 pc scales are very similar in Figure 19. With the same archives, Observation IV studied the core fields and the polarization within 100 - 200 pc (accumulation length) of each core, and showed a significant correlation (Figure 6).  This means that the structures of Galactic B-fields at the scale of cloud formation are preserved in the cores.

 \begin{figure*}[float]
 \epsscale{1.5}
\plotone{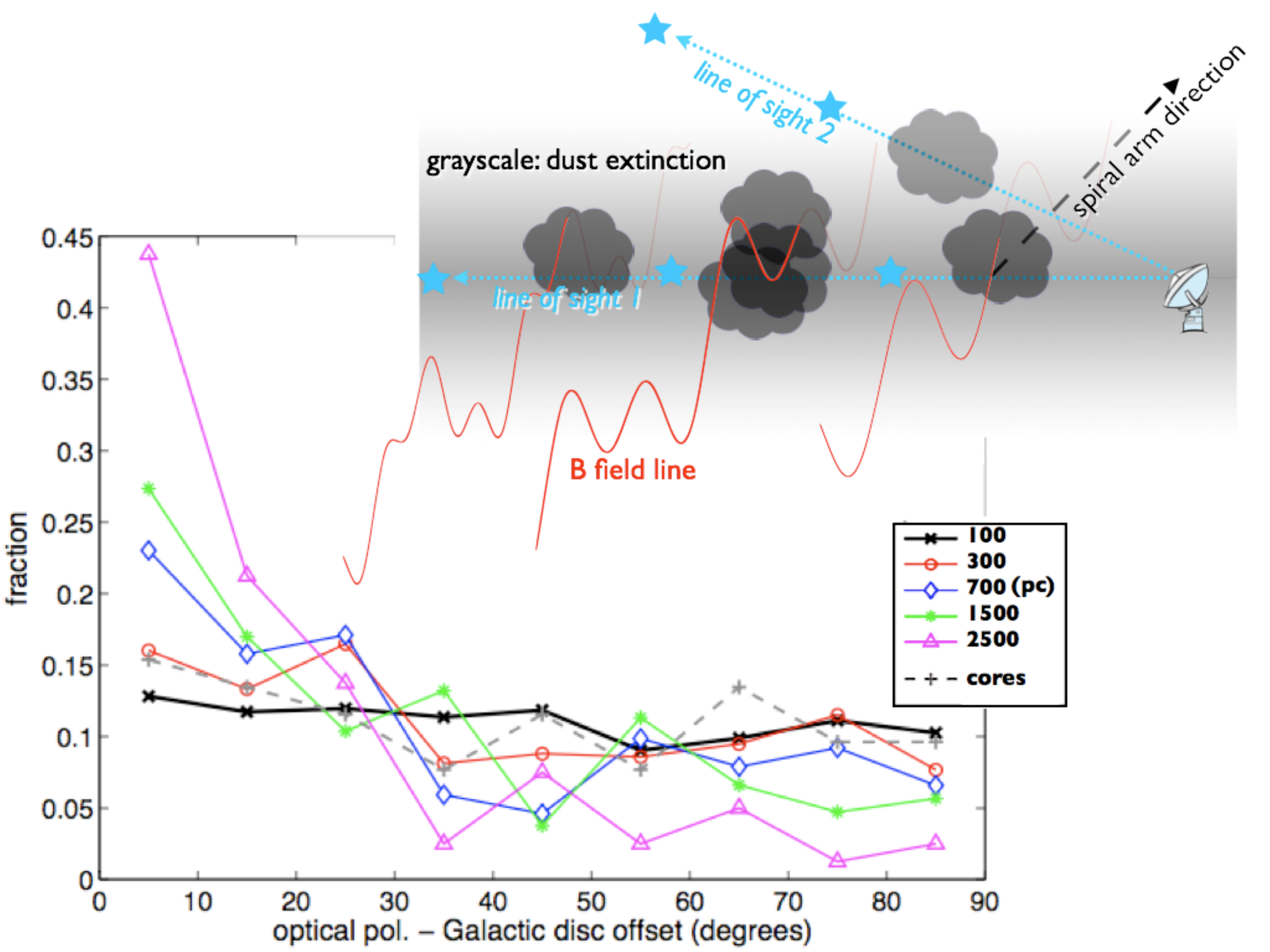}
 \caption{\small  Upper panel: An illustration of the fact that galactic B-fields (red lines) follow spiral arms (dark dashed arrow) and anchor clouds (Observation I and IV), but have rich structures perpendicular to the galactic disc. Compared to line of sight 1, line of sight 2 passes through less galactic mass other than one particular cloud. Stars with larger distance provide averaged field directions corresponding to larger scales.
Lower panel: We sample the stars with reliable polarization detections from 100-pc bins centered at distance 100, 300, 700, 1500, and 2500 pc.  The plots show that the Galactic B-field is more coherent at larger scales (above 700 pc), but is almost random at scales near 100 pc. Also plotted is the distribution of the B-fields from cloud cores (dashed line) at pc to sub-pc scales, probed with thermal dust emission ({\em Dotson et al.} 2010). The core field distribution is very similar to that of Galactic fields at 100-300 pc scales, the size of the accumulation length of a typical molecular cloud.
 }  
 \end{figure*}

\bigskip
\centerline{\textbf{ 7. SUMMARY }}
\bigskip

Recent B-field surveys and other related observations are discussed (Figure 1). 

Observation I (the correlation between cloud fields and spiral arms) and II (the constant ICM field strength) show that galactic B-fields are strong enough to hinder cloud rotation, channel gravitational contraction and imprint their directions onto molecular clouds. 
Only after the accumulated mass reaches a critical value defined by the galactic B-field strength ($\sim10\ \mu$G) can the cloud also contract in a direction perpendicular to the field lines and increase the field strength (Observation III). The cloud contraction density threshold can be reached by accumulation along B-fields; AD is not necessary for cloud contraction. However, observational dis/proof of the existence of AD is very challenging (Observation IX). Observation IV shows that the galactic field directions anchor into clouds all the way down to cloud cores. This B-field direction will make the ISM anisotropic in density profiles (Observations V, VI and VIII) and in velocity profiles of turbulence (Observation VII). The fact that cloud cores are flattened with the short axes oriented close the B-fields (Observation VIII) implies that, although cores might form under the super critical state (Observation III), they cannot be so super critical that the contraction becomes isotropic.

While a number of mechanisms have been proposed to explain how protostellar discs can overcome magnetic braking (see Chapter by Zhi-Yun Li et al.), observations of the B-field effects on disc formation are still at a very early stage. To our knowledge, two types of observations have been tried: (1) Probing the B field-turbulence decoupling scale -- two methods (Observation X and XI) agree on a value of several milli parsecs, which is comparable to the scale of disc formation. (2) Studying core field-outflow alignments -- assuming outflow directions as an indication of disc orientations (because outflow directions are easier to identify), Observation XII is aimed at studying whether core fields have any effect on disc orientations. However, we note that, like turbulent flows and gravitational contraction, outflows and disc rotation also interact with core B-fields. One cannot know for sure whether the outflows and/or disc rotation have destroyed the alignment or created fake alignments, especially on relatively small (100-1000AU) scales. Multiple-scale observations, similar to Observation IV, to determine whether B-field directions are changed in the outflows, discs, and vicinities, are critical for these kinds of experiments. 

At the end (section 6), we show that the observations used to criticise the strong-field scenario are not without their own difficulties upon close examination. ALMA should be able to further test the strong-field scenario (Observations IV-VIII) in regions with even higher densities.

\bigskip
\centerline{\textbf{ ACKNOWLEDGEMENT }}

HL appreciates the support from the Deutsche Forschungsgemeinschaft priority program 1573 (``Physics of the Interstellar Medium").
ZYL is supported in part by NASA NNX10AH30G, NNX14AB38G, and NSF AST1313083
\bigskip

\centerline\textbf{ REFERENCES}
\parskip=0pt
{\small
\baselineskip=11pt

\refs Allen, A., Li, Z.-Y. and Shu, F. (2003) {\em Astrophys. J., 599}, 363

\refs Andr\`e, Ph., Men'shchikov, A. and Bontemps, S. et al (2010) {\em Astron. Astrophys, 518}, 102 

\refs Arzoumanian, D., Andr\`e, Ph., Didelon, P. et al. (2011) {\em Astron. Astrophys, 529}, 6

\refs Attard, M., Houde, M., Novak, G., Li, H.-b., Vaillancourt, J., Dowell, D., Davidson, J. and Shinnaga, H. (2009) {\em Astrophys. J. 702}, 1584

\refs Benson, P. and Myers, P. (1989) {\em Astrophys. J. Supp., 71}, 89

\refs Bertram, E., Federrath, C., Banerjee, R. and Klessen, R. (2012) {\em Mon. Not. Roy. Astr. Soc., 420}, 3163

\refs Barranco, J. and Goodman, A. (1998) {\em Astrophys. J., 504}, 207

\refs Bierman, E., Matsumura, T. and Dowell, C. et al. (2011)  {\em Astrophys. J., 741}, 81

\refs Bourke, T.,  Myers, P.,  Robinson, G. and Hyland, A. (2001) {\em Astrophys. J., 554}, 916

\refs Chandrasekhar, S. and Fermi, E (1953) {\em Astrophys. J., 118}, 113

\refs Chapman, N., Goldsmith, P., Pineda, J., Clemens, D., Li, D. and Krco, M. (2011)  {\em Astrophys. J., 741}, 21
\refs Chapman, N., Davidson, J., Goldsmith, P., Houde, M., Kwon, W., Li, Z.-Y., Looney, L., Matthews, B., Matthews, T., Novak, G., Peng, R., Vaillancourt, J. and Volgenau, N. (2013) {\em Astrophys. J., 770}, 151

\refs Chitsazzadeh, S., Houde, M., Hildebrand, R. and Vaillancourt, J. (2012)  {\em Astrophys. J., 749}, 45

\refs Cho J. and Lazarian A. (2002) {\em Phys. Rev. Lett., 88}, 245001

\refs Cho J. and Vishniac E. T. (2000) {\em Astrophys. J., 539}, 273

\refs Cho J., Lazarian A. and Vishniac E. T. (2002) {\em Astrophys. J., 564}, 291

\refs Ciolek, G. and Mouschovias, T. (1993) {\em Astrophys. J., 418}, 774

\refs Clemens, D., Sarcia, D., Grabau, A., Tollestrup, E., Buie, M., Dunham, E. and Taylor, B. (2007) {\em Publications of the Astronomical Society of the Pacific, 119}, 1385

\refs 	Clemens, D., Pavel, M. and Cashman, L. (2012) {\em Astrophys. J. Supp., 200}, 21

\refs Collins, D., Kritsuk, A., Padoan, P., Li, H., Xu, H., Ustyugov, S. and Norman, M. (2012) {\em Astrophys. J., 750}, 13-18

\refs Cortes, P., Crutcher, R. and Watson, W. (2005) {\em Astrophys. J., 628}, 780

\refs Crutcher, R., Troland, T., Lazareff, B., Paubert, G. and Kaz\`{e}s, I. (1999)  {\em Astrophys. J. Let., 514}, 121

\refs Crutcher, R., Hakobian, N. and Troland, T. (2009) {\em Astrophys. J., 692}, 844-855

\refs Crutcher, R., Hakobian, N. and Troland, T. (2010) {\em Mon. Not. Roy. Astr. Soc., 402}, 64

\refs Crutcher, R., Nutter, D., Ward-Thompson, D. and Kirk, J. (2004)  {\em Astrophys. J., 600}, 279

\refs Crutcher, R. M., Wandelt, B., Heiles, C., Falgarone, E. and Troland, T. H. (2010) {\em Astrophys. J., 725}, 466-479

\refs Crutcher,  R. (2012) {\em Ann. Rev. Astron. Astrophys., 50}, 29-63

\refs Curran, R. and Chrysostomou, A. (2007), {\em Mon. Not. Roy. Astr. Soc., 382}, 699

\refs Davidson, J., Novak, G., Matthews, T., Matthews, B., Goldsmith, P, Chapman, N., Volgenau, N., Vaillancourt, J. and Attard, M. (2011)

\refs Dobashi, K. (2011) {\em PASJ,  63}, 1

\refs Dobbs, C. G (2008)  {\em Mon. Not. Roy. Astr. Soc., 391}, 844-858.

\refs Dotson, J. (1996) {\em Astrophys. J., 470}, 566 

\refs Dotson J. L., Vaillancourt J. E., Kirby L., Dowell C. D., Hildebrand R. H. and
Davidson J. A. (2010) {\em Astrophys. J. Supp., 186}, 406

\refs Dowell, D., Cook, B., Harper, D. et al. (2010) {\em Proceedings of the SPIE, 7735}, 213

\refs Elmegreen, B. (2007) {\em Astrophys. J., 668}, 1064

\refs Falceta--Gon\c{c}alves, D., Lazarian, A. and Kowal, G. (2008) {\em Astrophys. J., 679}, 537

\refs Fauvet, L., Mac\'ias-P\'erez, J., Jaffe, T., Banday, A., Desert, F. and Santos, D. (2012) {\em Astron. Astrophys, 540}, 122

\refs Fissel, L. et al. (2010) {\em Proceedings of the SPIE, 7741}, 9

\refs Frisch, U. (1995) {\em Turbulence: The Legacy of A. N. Kolmogorov} (Cambridge: Cambridge Univ. Press)

\refs Froebrich, D. and Rowles, J. (2010) {\em Mon. Not. Roy. Astr. Soc., 406}, 1350-1357

\refs Galli, D. and Shu, F. (1993) {\em Astrophys. J., 417}, 220

\refs Girart, J., Rao, R. and Marrone, D. {\em Science 313}, 812

\refs 	Goldreich, P. and Kylafis, N. (1981) {\em Astrophys. J., 243}, 75

\refs Goldreich P. and Sridhar H. (1995) {\em Astrophys. J., 438}, 763

\refs Goldreich P. and Sridhar H. (1997) {\em Astrophys. J., 485}, 680

\refs Goldsmith, P., Heyer, M., Narayanan, G., Snell, R., Li, D. and Brunt, C. (2008)  {\em Astrophys. J., 680}, 428

\refs Goodman, A., Bastien, P., Menard, F. and Myers, P. (1990) {\em Astrophys. J., 359}, 363

\refs Goodman, A., Pineda, J. and Schnee, S. (2009)  {\em Astrophys. J., 692}, 91

\refs Goodman, A., Barranco, J., Wilner, D. and Heyer, M. (1998) {\em Astrophys. J., 504}, 223

\refs Hall, J. (1951) {\em Astrophys. J., 56}, 40

\refs Han, J. L. and Zhang, J. S. (2007) {\em Astron. Astrophys, 464}, 609 

\refs Hartmann, L., Ballesteros-Paredes, J. and Bergin, E. (2001) {\em Astrophys. J., 562}, 852

\refs Heiderman, A., Evans, N., Allen, L., Huard, T. and Heyer, M. (2010) {\em Astrophys. J., 723}, 1019-1037

\refs Heiles, C. (2000) {\em Astron. J., 119}, 923

\refs Heiles, C., Goodman, A., McKee, C. and Zweibel, E. (1993) Protostars and planets III, 279

\refs Heiles, C and Troland, T. (2005) {\em Astrophys. J., 624}, 773

\refs Hennebelle, P., Commercon, B., Joos, M., Klessen, R. S., Krumholz, M., Tan, J. C. and
Teyssier, R. (2011) {\em Astron. Astrophys, 528}, 72

\refs Henning, Th., Linz, H., Krause, O., Ragan, S., Beuther, H., Launhardt, R., Nielbock, M. and 
Vasyunina, T. (2010) {\em Astron. Astrophys, 518}, 95 

\refs Heyer M., Gong H., Ostriker E. and Brunt C. (2008)  {\em Astrophys. J., 680}, 420

\refs Heyer, M. H. and Brunt, C. M. (2012) {\em Mon. Not. Roy. Astr. Soc., 420}, 1562-1569

\refs Hezareh, T., Houde, M., McCoey, C. and Li, H. (2010) {\em Astrophys. J., 720}, 603

\refs Hildebrand, R., Kirby, L., Dotson, J., Houde, M. and Vaillancourt, J. (2009) {\em Astrophys. J., 696}, 567

\refs Hill, T., Motte, F. and Didelon, P., et al. (2011)  {\em Astron. Astrophys, 533}, A94 

\refs Hiltner, W. (1951) (1951) {\em Astrophys. J., 114}, 241

\refs 	Houde, M., Bastien, P., Peng, R., Phillips, T. and Yoshida, H. (2000 a) {\em Astrophys. J., 537}, 245

\refs 	Houde, M., Peng, R., Phillips, T., Bastien, P. and Yoshida, H. (2000 b) {\em Astrophys. J., 536}, 857

\refs Houde, M., Bastien, P., Dotson, J., Dowell, D.,  Hildebrand, R., Peng, R., Phillips, T., Vaillancourt, J. and Yoshida, H. (2002) {\em Astrophys. J., 569}, 803

\refs 	Houde, M., Vaillancourt, J., Hildebrand, R., Chitsazzadeh, S. and Kirby, L. (2009)  {\em Astrophys. J., 706}, 1504

\refs Houde, M., Rao, R., Vaillancourt, J. and Hildebrand, R. (2011) {\em Astrophys. J., 733}, 109

\refs Houde, M., Fletcher, A., Beck, R., Hildebrand, R., Vaillancourt, J. and Stil, J. (2013) {\em Astrophys. J., 766}, 49

\refs Hull, C., Plambeck, R., Bolatto, A. et al. (2013) {\em Astrophys. J., 768}, 159

\refs 	Jones, C., Basu, S. and Dubinski, J. (2001) {\em Astrophys. J., 551}, 387

\refs Joos, M., Hennebelle, P. and Ciardi, A. (2012) {\em Astron. Astrophys, 543}, 128

\refs Kainulainen, J., Beuther, H., Henning, T. and  Plume, R. (2009) {\em Astron. Astrophys, 508}, 35-39

\refs Kandori, R. et al. (2006) {\em Proceedings of the SPIE, 6269}, 159

\refs Lada, C.. Lombardi, M. and Alves, J. (2010) {\em Astrophys. J., 724}, 687 

\refs Lazarian, A., Esquivel, A. and Crutcher, R. (2012) {\em Astrophys. J., 757}, 154

\refs Li, H.-b., Fang, M., Henning, T. and Kainulainen, J. (2013) {\em Mon. Not. Roy. Astr. Soc., 436}, 3707 

\refs Li, H.-b. and Henning, T. (2011) {\em Nature, 479}, 499-501

\refs Li, H.-b., Blundell, R., Hedden, A., Kawamura, J., Paine, S. and Tong, E. (2011) {\em Mon. Not. Roy. Astr. Soc., 411}, 2067-2075

\refs Li H.-b., Houde M., Lai S.-p. and Sridharan T. K. (2010) {\em Astrophys. J., 718}, 905

\refs Li, H.-b., Dowell, C. D., Goodman, A., Hildebrand, R. and Novak, G. (2009)  {\em Astrophys. J., 704}, 891 

\refs Li, H.-b. and Houde M. (2008a)  {\em Astrophys. J., 677} 1151

\refs Li, H., Dowell, D., Kirby, L., Novak, G. and Vaillancourt, J. (2008b) {\em Applied Optics 47} 422

\refs Li, H.-b., Griffin, G., Krejny, M., Novak, G., Loewenstein, R., Newcomb, M., Calisse, P. G. and Chuss, D. (2006) {\em Astrophys. J., 648}, 340

\refs Li, P.- S., McKee, C., Klein, R. and Fisher, R. (2008a) {\em Astrophys. J., 684}, 380-394

\refs Li, P.- S., McKee, C. and Klein, R. (2012)  {\em Astrophys. J., 744}, 73

\refs Li, Z.-Y., Krasnopolsky, R. and Shang, H. (2011)  {\em Astrophys. J., 738}, 180

\refs Lunttila, T., Padoan, P., Juvela, M. and Nordlund, Ake (2008) {\em Astrophys. J., 686}, 91-94

\refs Machida, M., Inutsuka, S.-I. and Matsumoto, T. (2011) {\em PASJ, 63}, 555

\refs Matthews, B., McPhee, C., Fissel, L., and Curran, R. (2009), {\em Astrophys. J. Supp., 182}, 143-204

\refs Matthews, T. et al. arxiv.org/abs/1307.5853, submitted to {\em Astrophys. J.}

\refs McKee, C., Zweibel, E., Goodman, A. and Heiles, C. (1993) Protostars and planets III, 327

\refs Mestel, L. and Spitzer, L., Jr. (1956) {\em Mon. Not. Roy. Astr. Soc., 116}, 503

\refs Mizuno A., Onishi T., Yonekura Y., Nagahama T. and Ogawa H. et al. (1995)  {\em Astron. Astrophys, 445}, 161

\refs Molinari, S., Swinyard, B. and Bally, J., et al. (2010) {\em Astron. Astrophys, 518}, 100 

\refs Maron J. and Goldreich P. (2001) {\em Astrophys. J., 554}, 1175

\refs Mouschovias, T. (1976)  {\em Astrophys. J., 207}, 141

\refs Mouschovias, T. and Paleologou, E. (1979) {\em Astrophys. J., 230}, 204

\refs Mouschovias, T. and Tassis, K. (2010) {\em Mon. Not. Roy. Astr. Soc., 409}, 801-807

\refs Mouschovias, T. and Tassis, K. (2009) {\em Mon. Not. Roy. Astr. Soc., 400}, 15

\refs Murillo, N. and Lai, S.-P. (2013)  {\em Astrophys. J., 764}, 15

\refs Myers, P. (2009) {\em Astrophys. J., 700}, 1609

\refs Nagahama, T., Mizuno, A., Ogawa, H. and Fukui, Y. (1998) {\em Astron. J., 116}, 336

\refs Nagai, T., Inutsuka, S.-I.; and Miyama, S. (1998) {\em Astrophys. J., 506}, 306

\refs Nakamura, F. and Li, Z. (2008) {\em Astron. J., 687}, 354

\refs Nakano, T. and Nakamura, T. (1978) {\em PASJ, 30}, 681

\refs Nordlund, A. K. and Padoan, P. (1999) {\em Interstellar Turbulence}, Proceedings of the 2nd Guillermo Haro Conference. Edited by Jose Franco and Alberto Carraminana. Cambridge University Press, 218-223 

\refs Novak, G. (2006) {\em SPIE Newsroom} DOI: 10.1117/2.1200609.0373.

\refs Padoan, P., Jimenez, R., Juvela, M. and Nordlund, �. (2004), {\em Astrophys. J., 604}, 49

\refs Padoan, P., Juvela, M., Goodman, A. and Nordlund, A. (2001)  {\em Astrophys. J., 553}, 227

\refs Palmeirim, P., Andr\'e, Ph., Kirk, J. et al. (2013) {\em Astron. Astrophys, 550}, 38

\refs Pascale, E., Ade, P., Angile, F. et al. (2012) {\em Proceedings of the SPIE, 8444}, 15

\refs Peretto, N.; Andr\'e, Ph.; Konyves, V. et al. (2012)  {\em Astron. Astrophys, 541}, 63

\refs Price, D. and Bate, M. (2008)  {\em Mon. Not. Roy. Astr. Soc., 385}, 1820 

\refs Ragan, S. et al. (2012) 2012arXiv1207.6518R

\refs 	Scandariato, G., Robberto, M., Pagano, I. and Hillenbrand, L. (2011) {\em Astron. Astrophys, 533}, 38

\refs Schneider, N., Bontemps, S., Simon, R., Ossenkopf, V., Federrath, C., Klessen, R., Motte, F., Andr\'e, Ph., Stutzki, J. and Brunt, C (2011) {\em Astron. Astrophys, 529}, 1

\refs Schneider, S.  and Elmegreen, B. (1979). {\em Astrophys. J. Supp., 41}, 87

\refs Schlegel, D., Finkbeiner, D. and Davis, M. (1998) {\em Astrophys. J., 50}, 525

\refs Seifried, D., Banerjee, R., Pudritz, R. and Klessen, R. (2012)  {\em Mon. Not. Roy. Astr. Soc., 423}, 40

\refs Shetty, R. and Ostriker, E. (2006) {\em Astrophys. J., 647}, 997-1017

\refs Shu, F., Allen, A., Shang, H., Ostriker, E. and Li, Z.-Y. (1999) The Origin of Stars and Planetary Systems. Edited by Charles J. Lada and Nikolaos D. Kylafis. Kluwer Academic Publishers, 193 

\refs Stephens, I., Looney, L., Dowell, C., Vaillancourt, J. and Tassis, K. (2011) {\em Astrophys. J., 728}, 99

\refs Stone, J., Ostriker, E. and Gammie, C. (1998) {\em Astrophys. J., 508}, 99 

\refs Tassis K., Dowell C. D., Hildebrand R. H., Kirby L. and Vaillancourt J. E.
(2009) {\em Mon. Not. Roy. Astr. Soc., 399}, 1681

\refs 	Tilley, D. and Balsara, D. (2010) {\em Astrophys. J., 406}, 1201

\refs  Tilley, D. and Balsara, D. (2011) {\em Mon. Not. Roy. Astr. Soc., 415}, 3681

\refs Tobin, J., Hartmann, L., Chiang, H.-F., Wilner, D., Looney, L., Loinard, L., Calvet, N. and D'Alessio, P. (2012) {\em Nature, 942}, 83

\refs Tomisaka, K. (2011) {\em Publications of the Astronomical Society of Japan, 63}, 147

\refs Troland, T. and Crutcher, R. (2008) {\em Astrophys. J., 680}, 457, 465
 
\refs Vestuto J. G., Ostriker E. C. and Stone J. M. (2003) {\em Astrophys. J., 590}, 858

\refs Vlemmings, W., Torres, R. and Dodson, R. (2011) {\em Astron. Astrophys, 529}, 95

\refs Williams, J., Blitz, L. and McKee, C. (2000) {\em Protostars and Planets IV}, ed. A. Mannings, P. Boss, and S. S. Russell (Tucson, AZ: Univ. Arizona Press), 97

\refs Wolf, S., Launhardt, R. and Henning, T. (2003) {\em Astrophys. J., 592}, 233

\refs Vaillancourt, J., Dowell, D., Hildebrand, R., Kirby, L., Krejny, M., Li, H.-b., Novak, G., Houde, M., Shinnaga, H. and Attard, M. (2008) {\em Astrophys. J., 679}, 25

\refs Zweibel, E. (2002)  {\em Astrophys. J., 567}, 962

\end{document}